\documentclass[12pt]{iopart}
\begin{document}

\title[GLASS TRANSITION IN SYSTEMS WITHOUT STATIC CORRELATIONS]
{GLASS TRANSITION IN SYSTEMS WITHOUT STATIC CORRELATIONS:
A MICROSCOPIC THEORY}

\author{R. Schilling$^{1}$ and G. Szamel$^{1,2}$}

\address{$^1$ Institut f\"ur Physik, Johannes Gutenberg-Universit\"at
Mainz, D-55099
Mainz, Germany}

\address{$^2$ Department of Chemistry, Colorado State University, Ft.
Collins, CO 80523, USA}

\begin{abstract}
We present a first step toward a microscopic theory for the glass
transition in systems with trivial static correlations. As an
example we have chosen $N$ infinitely thin hard rods with length
$L$, fixed with their centers on a periodic lattice with lattice
constant $a$. Starting from the $N$-rod Smoluchowski equation we
derive a coupled set of equations for fluctuations of reduced
$k$-rod densities. We approximate the influence of the surrounding
rods onto the dynamics of a pair of rods by introduction of an
effective rotational diffusion tensor $\mathbf{D} (\Omega_1,
\Omega_2)$ and in this way we obtain a self-consistent equation
for $\mathbf{D}$. This equation exhibits a feedback mechanism
leading to a slowing down of the relaxation. It involves as an
input the Laplace transform $\upsilon_0(l/r)$ at $z=0$, $l=L/a$,
of a torque-torque correlator of an isolated pair of rods with
distance $R=ar$. Our
equation predicts the existence of a continuous
ergodicity-breaking transition at a critical length $l_c=L_c/a$.
To estimate the critical length we perform an approximate
analytical calculation of $\upsilon_0(l/r)$ based on a
variational approach and obtain $l_c^{\mathrm{var}} \cong 5.68$,
4.84 and 3.96 for an sc, bcc and fcc lattice. We also evaluate
$\upsilon_0(l/r)$ numerically exactly from a two-rod
simulation. The latter calculation leads to $l_c ^{\mathrm{num}}
\cong 3.45$, 2.78 and 2.20 for the corresponding lattices. Close
to $l_c$ the rotational diffusion constant decreases as $D(l) \sim
(l_c - l)^\gamma$ with $\gamma=1$ and a diverging time scale
$t_\epsilon \sim |l_c - l|^{-\delta}$, $\delta=2$, appears. On this
time scale the $t$-- and
$l$--dependence of the 1-rod density is determined by a master
function depending only on $t/t_\epsilon$. In contrast to
present microscopic theories our approach predicts a glass
transition despite the absence of any static correlations.
\end{abstract}


\submitto{\JPCM}

\maketitle

\section{INTRODUCTION}

In the last two decades huge theoretical efforts were made in
order to describe the glass transition on a \textit{microscopic}
level in systems \textit{without} quenched disorder. Microscopic
means that, \emph{e.g.}, the glass transition temperature and the
dynamical properties in its vicinity can be calculated from
equations which use as an input the knowledge of the interactions
between microscopic constituents of the system (atoms or
molecules; hereafter we use the term particles). Most of this
activity has been devoted to \textit{structural glasses}
\cite{1,2,3,4}. Although experiments \cite{5a,6,7,5} have
demonstrated that plastic crystals can exhibit an
\textit{orientational} glass transition which has many features in
common with the structural glass transition there seems to be no
analytical theory for such systems. \footnote[1]{The microscopic
theory developed by Michel and coworkers \cite{8,9} is valid for the
orientational glass transition of mixed crystals, \emph{i.e.}
crystals with \textit{quenched} disorder.} On the other hand
Molecular Dynamics (MD) simulations \cite{10,11,12} have confirmed
the similarity between glassy behavior of plastic crystals and
that of supercooled liquids.

The first successful microscopic theoretical approach to glassy
dynamics in supercooled liquids came from so-called mode
coupling theory (MCT) \cite{1,2}. This theory was first suggested
and mainly worked out in details by G\"otze and his
coworkers. For reviews the reader may consult refs.
\cite{13,14,15}. MCT derives a closed set of equations for
time-dependent correlators, like the intermediate scattering
function $S(q,t)$. This set uses as an
input the corresponding \textit{static} correlations, \emph{e.g.}
$S(q)$, which
depend only on thermodynamic variables like temperature $T$, density
$n$, \textit{etc.}
The static correlations can be calculated from the microscopic
interactions by use of either analytical or numerical tools
\cite{16}. By
decreasing $T$ (or increasing $n$) the peaks of $S(q)$ grow.
If their magnitude becomes large enough a transition from an
ergodic to a non-ergodic phase occurs at a critical temperature
$T_c$ (or density $n_c$). This \textit{dynamic} transition is
interpreted as an ideal glass transition. MCT makes several
nontrivial predictions for, \emph{e.g.}, time dependence of
correlators \cite{13,14,15} which have been successfully
tested (see \cite{17,18,19,Kob} for reviews).

The second microscopic theory of the glass transition has been
developed by M\'ezard and Parisi \cite{3}. Their first principle theory
combines the use of the replica idea known from the spin glass
theory and of the liquid state theory techniques. It is related
to an earlier density functional approach
\cite{4}. M\'ezard and Parisi's theory
starts from the microscopic Hamiltonian for $m$
identical replicas of the system
to which a symmetry breaking field is added. The
free energy of the replicated system is then calculated and the
configurational entropy of the original system is extracted from
it. The theory predicts the existence of a critical temperature
$T_f$ (which is below $T_c$ of MCT) at which the configurational
entropy vanishes. This is then interpreted as an ideal
\textit{static} glass transition.

Besides these two microscopic theories, let us mention one
other attempt to describe the glass transition that does not resort to
microscopic potentials: based on several
\textit{qualitative} assumptions Edwards and Vilgis \cite{23} obtained
the
dependence of the self-diffusion constant $D$ for a system of hard rods
on physical control parameter $x$, like, \emph{e.g.},
concentration, diameter, or length of the rods. A dynamical
glass transition occurs when $D$ becomes zero. A mean field
treatment results in $D \sim x_c-x$ and an improved approach leads
to the Vogel-Fulcher law $D \sim \exp [A/(x_c-x)]$. Close to the
critical point $x_c$ a power law dependence of $D$ on $(x_c-x)$ is found
with a non-trivial exponent $\gamma=7/6$.

Both above described microscopic theories have at least one
feature in common: non-trivial equilibrium correlations. These
correlations are of primary importance for both theories. For
example, MCT requires that the vertices which enter into the
self-consistent equations for the correlation functions must
depend on at least one control parameter. Recently, it has been
shown that a colloidal suspension with hard-core plus attractive Yukawa
potential can undergo a gel transition \cite{24} in a special
limit: $\phi \rightarrow 0$ and $K \rightarrow \infty$ such that
$\Gamma =K ^2 \phi / b= \textrm{const}$. Here $\phi$ is the
volume fraction, $K$ the interaction strength and $b$ a measure of
the inverse range of interaction. In that limit the direct
correlation function $c(q)$ is proportional to $K$ and becomes
independent of $\phi$ \cite{25}.  This implies that the static
correlations vanish, \emph{i.e.} $S(q) \rightarrow 1$. But since
the vertices are proportional to $\phi$ and involve a bilinear
product of $c(q)$ they become proportional to $\Gamma$.
Accordingly, despite the vanishing of the static correlations
there is still an ergodicity breaking transition driven by the
increase of $\Gamma$ \cite{24,25,26}. Of course, this type of
behavior is not generic in the $\phi-K$ phase space. In addition,
whether the mode coupling approximation remains reasonable must be
still checked.

One may now ask the following question: Can there be a
\textit{generic} glass transition if and only if there exist
nontrivial static correlations? To find an answer let us consider
a liquid of hard rods. This system was used as a paradigm in
ref. \cite{23}. If the rods thickness $d$ is finite the static
correlations are non-trivial and a glass transition will be predicted
by either theory.
However, in the limit of \textit{infinitely thin} hard rods,
\textit{i. e.} $d=0$,
with \textit{finite} length $L$ and \textit{finite} concentration
$c$, the static properties become trivial. Note that in this limit
there is no transition to a nematic phase. Also, according to either
of the two microscopic theories there is no glass transition!
In the case of MCT the vertices vanish.
The absence of a glass transition is also consistent with
a microscopic theory worked out by one of the authors and his
coworkers \cite{27,28}. There, it has been shown that for infinitely
thin hard rods with randomly frozen orientations
translational diffusion never ceases if the diffusion constant
along the rod's direction is nonzero.

Now suppose we fix the infinitely thin hard rods with their
centers on the sites of a three-dimensional periodic lattice. Such
a model has already been suggested earlier \cite{23,29}. It can be
used to describe the steric hindrance of the orientational degrees
of freedom in plastic crystals \cite{6,7,5,10,11,12}. Again, both
MCT and the replica theory exclude a glass transition, due to the
lack of static correlations. The fact that MCT in its present form
does not predict a glass transition was already stressed in ref.
\cite{30}. Nevertheless, computer simulations \cite{30,31} 
showed glassy behavior and suggested 
a \textit{dynamical} ergodicity-breaking transition 
at a critical length $L_c$.
\footnote{Not that different cubic lattices were used in refs. \cite{30}
and \cite{31}; moreover the authors of ref. \cite{31} fixed the rods
with one of their endpoints at the lattice sites.} This simple
example makes clear that non-trivial static behavior is
\textit{not} necessary for a glass transition. It is the main goal
of our paper to present a microscopic approach to the glass
transition for systems without static correlations.

A short account of our approach for the case of a simple cubic
lattice was given in ref. \cite{32}. In the present paper we will
discuss our model and the theoretical framework in more detail and
will extend the results to bcc and fcc lattices. Furthermore, we
will investigate the type of glass transition and the long time
dynamics in its vicinity.

The outline of our paper is as follows. In the next section we
will present and discuss the model. The third
section contains the analytical theory used to describe the glass
transition. The results are given in the fourth section and the
final section includes a summary and some conclusions.

\section{MODEL}
We consider $N$ hard, infinitely thin rods (\emph{i.e.} diameter
$d=0$) with length $L$ and with their centers fixed at the lattice
sites of a \textit{periodic} lattice with lattice constant $a$.
The dimensionless control parameter is the reduced rod length, $l=
L/a$. In the following we will restrict ourselves to a simple
cubic lattice. Its unit cell is depicted in Fig.~\ref{fig1}. All
qualitative conclusions which will be drawn in this paper remain 
valid for \textit{all} periodic lattices.

It is obvious that for $l \leq 1$ each rod can rotate freely:
there is no steric hindrance. For $l$ above one, collisions
between a rod and adjacent ones appear. If $1 \leq l \leq
\sqrt{3}$ a given rod interacts only with the six nearest
neighbors. Further increase of $l$ will introduce collisions
between next-, third-, fourth-, \textit{etc.} 
nearest neighbors, generating
a strong increase of steric hindrance. Since we have chosen
infinitely thin rods there is no equilibrium transition to an
orientationally ordered phase. On the other hand, due to the rise
of steric hindrance with $l$ we expect an orientational glass
transition as detected in ref. \cite{30} and \cite{31}. Although
steric hindrance already exists for $l > 1$, it seems that no
ergodicity-breaking transition can occur for $l < 2$. Let us
consider the following simple argument. 
If $l > 2$ then the nearest neighbor rods of a tagged rod (rod 1)
can ``overlap'' with, \emph{e.g.}, the top of the tagged rod 
(see Fig. 2). This induces a ``cage'' for the tagged rod.
Note that this argument is not rigorous, because for $l < 2$ there 
might exist other configurations similar to the one shown in
Fig.~\ref{fig2} which could block the tagged rod's rotation. But
the value $l=2$ also plays a role from a different point of view.
Consider two neighboring rods with their centers along the
$z$-axis. Their orientations can be described by
$\Omega_i=(\theta_i, \phi_i)$ and $\Omega_j=(\theta_j, \phi_j)$.
At a collision it is either $\phi_j=\phi_i^{\pm}$ (see
Fig.~\ref{fig3} a) or $\phi_j=\phi_i^{\pm} + \pi$ (see
Fig.~\ref{fig3} b), for fixed $\theta_i$, $\theta_j$ and $\phi_i$.
The notation $\phi_i^+$ and $\phi_i^-$ means that rod $j$ is
behind and in front of rod $i$, respectively (see Figure
~\ref{fig3}). Now, it is easy to prove that for \textit{arbitrary}
$\theta_i$ and $\theta_j$ there exists only \textit{one} contact
(either at $\phi_i^{\pm}$ \textit{or} at  $ \phi_i^{\pm} + \pi$) 
for $l < 2$ (see Fig. ~\ref{fig3} a,b). In that case
$\phi_j$ can be varied freely by $2 \pi$. But for $l > 2$ there is
a finite range for both angles $\theta_i$ and $\theta_j$ such that
\textit{two} contacts (at $\phi_i^{\pm} $  \textit{and} $
\phi_i^{\pm} + \pi )$ exist (see Fig.~\ref{fig3} c). Therefore the
``phase space'' $[0, 2 \pi]$ for $\phi_j - \phi_i$ is decomposed
into two ``ergodic'' components $[0, \pi]$ and $[\pi, 2 \pi]$ for
$\theta_i$ and $\theta_j$ kept fixed. Transitions between both
components are forbidden. If the rods were fixed at the lattice
sites with one of their end points (like in ref. \cite{31}) then
there is only one contact, for \textit{all} $l > 1$.

The condition for a collision can be quantified as follows
\cite{33}. Let $\mathbf{u}_i$ and $\mathbf{u}_j$ be the unit
vectors along rod $i$ and $j$ and $\mathbf{r}_{ij}$ the vector
connecting the center of rod $j$ with that of rod $i$ (see
Fig.~\ref{fig4}). $\mathbf{u}_i$ and $\mathbf{u}_j$ define a
plane. Let $\mathbf{r}^{\bot}_{ij} $ be the component of
$\mathbf{r}_{ij}$ perpendicular to that plane and $r_{ij}^{\bot}=
|\mathbf{r}^{\bot}_{ij}|$. Denote the distance of the point of contact to
the center of rod $i$ and $j$ by, respectively, $s_{ij}$ and
$s_{ji}$ (Fig.~\ref{fig4}). Then a collision occurs if:

\begin{equation} \label{eq1}
r_{ij}^{\bot} = 0^+ \quad , \quad |s_{ij}| < L/2 \quad , \quad
|s_{ji}| < L/2
\end{equation}
where $0^+$ stands for limit $d \rightarrow 0$. Simple algebra
yields:

\begin{equation} \label{eq2}
s_{ij}= -[(\mathbf{u}_i \cdot\mathbf{r}_{ij})- (\mathbf{u}_i
\cdot\mathbf{u}_{j}) (\mathbf{u}_j \cdot\mathbf{r}_{ij})] /
[1-(\mathbf{u}_i
\cdot\mathbf{u}_{j})^2]
\end{equation}

\begin{equation} \label{eq3} s_{ji}= [(\mathbf{u}_j
\cdot\mathbf{r}_{ij})- (\mathbf{u}_i \cdot\mathbf{u}_{j}) (\mathbf{u}_i
\cdot\mathbf{r}_{ij})] / [1-(\mathbf{u}_i \cdot\mathbf{u}_{j})^2]
\end{equation}
Note (i) that $s_{ij}$ and $s_{ji}$ can be negative; in that case
the contact points into the direction of $- \mathbf{u}_i$ and
$-\mathbf{u}_{j}$, respectively and (ii) that $\mathbf{r}_{ij}$ is
time-independent and is always equal to a lattice vector
$\mathbf{R}_{ij}= \mathbf{R}_i - \mathbf{R}_j$.

The configurational part of the partition function is given by

\begin{equation} \label{eq4} 
Z_c (T, N; l) = \int \prod\limits_{i=1}^{N} d \Omega_i \,\,e^{-
\beta V(\Omega_1, \cdots, \Omega_N)}= (4 \pi)^N \quad , \quad
\beta=(k_BT)^{-1}
\end{equation}
since the hard rod interaction potential $V( \Omega_1, \cdots,
\Omega_N)$ vanishes almost everywhere. The corresponding contribution
$-k_BT
N\ln 4 \pi$ to the free energy is analytic in temperature $T$ and
does not depend on $l$. Hence, there is no equilibrium phase
transition. The equilibrium pair distribution function
$g_{ij}^{(2)} (\Omega_i, \Omega_j)$ is related to the probability
to find a rod at $\mathbf{R}_i$ and $\mathbf{R}_j (\neq
\mathbf{R}_i)$ with orientations $\Omega_i$ and $\Omega_j$,
respectively. It is given by:

\begin{equation} \label{eq5}
g_{ij}^{(2)} \big(\Omega_i, \Omega_j\big)= 1 - \theta \left(0^+ -
r^\bot_{ij}\right) \theta \left(\frac{L}{2} - |s_{ij}|\right)
\theta \left(\frac{L} {2} - |s_{ji}|\right)
\end{equation}
where the Heaviside functions are nonzero at a collision, which
happens if conditions Eq.~(\ref{eq1}) are fulfilled. Since a
collision is non-generic, $g_{ij}^{(2)} (\Omega_i,
\Omega_j)=1$ for almost all $\Omega _i$ and $\Omega_j$. Hence
there are no static correlations, except on a set of measure zero.

\section{KINETIC THEORY}

In this section we will describe the dynamics for our model and
the approximations leading to a closed set of equations. We
assume that the microscopic $N$-rod dynamics of the system is Brownian
rather than Newtonian. Previous studies of systems with nontrivial static
correlations indicated that glassy
dynamics should not depend on that choice \cite{34,35}.

The starting point of the theory
is the so-called generalized Smoluchowski equation
for the $N$-rod probability density $P_N(\Omega_1, \cdots,
\Omega_N ;t)$ \cite{33}

\begin{equation} \label{eq6}
\frac{\partial} {\partial t} \, P_N (\Omega_1, \cdots, \Omega_N ;
t)= D_0 \, \sum\limits_{n=1}^{N} \, \nabla_n \cdot [ \nabla_n -
\mathbf{T}_n (\Omega_1, \cdots, \Omega_N)] P_N (\Omega_1, \cdots,
\Omega_N ; t)
\end{equation}
where $D_0$ is the \textit{bare} rotational diffusion constant,
$\nabla_n \equiv \nabla_{\Omega_n}\equiv
\mathbf{u}_n\times\nabla_{\mathbf{u}_n}$, and

\begin{equation} \label{eq7}
\mathbf{T}_n (\Omega_1, \cdots, \Omega_N) = \sum\limits_{j \neq n}
\, \mathbf{T}_{jn} (\Omega_j, \Omega_n).
\end{equation}
$\mathbf{T}_{ij} (\Omega_i, \Omega_j)$ describes the singular
torque when two rods at sites $i$ and $j$ with vector distance
$\mathbf{r}_{ij}$ collide:

\begin{equation} \label{eq8}
\mathbf{T}_{ij} \big(\Omega_i, \Omega_j\big)= s_{ij}
\big(\mathbf{u}_i \times \widehat\mathbf{r}_{ij}^{\bot}\big)
\delta (r_{ij}^\bot - 0^+) \theta \left(\frac{L}{2}-|s_{ij}|\right)
\theta \left(\frac{L}{2} - |s_{ji}|\right)
\end{equation}
where $\widehat\mathbf{r}_{ij}^{\bot}= \mathbf{r}_{ij}^\bot /
r_{ij}^\bot$. Using Eq.~(\ref{eq5}) it is easy to prove that

\begin{equation} \label {eq9}
\nabla_i \, g_{ij}^{(2)} \, (\Omega_i, \Omega_j) = \mathbf{T}_{ij}
(\Omega_i, \Omega_j) \, g^{(2)}_{ij} (\Omega_i, \Omega_j).
\end{equation}
Eq.~(\ref{eq9}) describes the noncrossability condition.
Note that $\mathbf{T}_{ij} (\Omega_i, \Omega_j)$ differs slightly
from $\mathbf{T}_{ij} (\Omega_i, \Omega_j)$ used in ref. \cite{33}
since there $\nabla_{u_i}$ was used instead of $\nabla_{\Omega_i}$.
Also, a similar generalized Smoluchowski equation had been used
before to describe a liquid of infinitely thin hard rods with randomly
frozen orientations \cite{27,28}.

In a next step we present equations of motion for the reduced
$k$-rod density

\begin{equation} \label{eq10}
\rho^{(k)}_{n_1 \cdots n_k} (\Omega_1, \cdots, \Omega_k ; t)= \,
\int \, \prod\limits_{n \neq n_1, \cdots, n_k} \,\,\, d \Omega_n \,P_N
(\Omega_1, \cdots, \Omega_ N ; t).
\end{equation}
These equations form an infinite, coupled hierarchy \cite{36} of
the following form

\begin{eqnarray} \label{eq11}
\frac{\partial} {\partial t} \, \rho_{n_1 \cdots
n_k}^{(k)}(\Omega_1, \cdots , \Omega_k ; t) = \mathcal{L}^{(k)}
\rho^{(k)}_{n_1 \cdots n_k} (\Omega_1, \cdots, \Omega_k ;
t)- \nonumber \\
- D_0 \sum\limits_{n \in I_k}  \sum\limits_{n_{k+1} {\not\in} I_k}
\, \nabla_n \cdot \int \, d \Omega_{k+1} \,\, \mathbf{T}_{n
n_{k+1}}
(\Omega_n, \Omega_{k+1} )\nonumber \\ \times
\rho_{n_1 \cdots n_{k+1}}^{(k +1)} (\Omega_1, \cdots, \Omega_{k+1}
; t)
\end{eqnarray}
with the $k$-rod Smoluchowski operator,

\begin{equation} \label{eq12}
\mathcal{L}^{(k)}= D_0 \, \sum\limits_{n \in I_k} \,\, \nabla_n
\cdot [\nabla_n - \sum\limits_{{n' \in I_k} \atop{n'(\neq n)}} \,
\mathbf{T}_{nn'} (\Omega_n, \Omega_{n'})].
\end{equation}
and $I_k= \{n_1, \cdots, n_k\}$. As initial conditions we choose:

\begin{equation} \label{eq13}
\rho^{(k)}_{n_1 \cdots n_k} \, \, (\Omega_1, \cdots, \Omega_k ; 0)
= \frac{1} {(4 \pi)^{k-1}} \,\, \delta (\Omega_1 | \Omega_0) \,
g^{(k)}_{n_1 \cdots n_k} (\Omega_1, \cdots, \Omega_k),
\end{equation}
where $\delta (\Omega| \Omega')= \sin \theta \delta
(\theta-\theta') \, \delta (\phi-\phi')$.
Here, $g^{(k)}_{n_1 \cdots n_k} (\Omega_1, \cdots , \Omega_k)$ is
the equilibrium $k$-rod distribution function which is equal to one
almost everywhere.

To proceed we introduce the fluctuations of the $k$-rod density
for $k \geq 2$

\begin{eqnarray} \label{eq14}
\delta \rho ^{(k)}_{{n_1} \cdots n_k} (\Omega_1, \cdots, \Omega_k;
t) = \rho^{(k)}_{n_1 \cdots n_k } (\Omega_1, \cdots, \Omega_k ;
t) \nonumber\\
 - \frac{1}{{( 4 \pi)}^{k-1}} \, g^{(k)}_{n_1 \cdots
n_k} (\Omega_1, \cdots, \Omega_k) \, \rho^{(1)} _{n_1} (\Omega_1 ;
t).
\end{eqnarray}
These fluctuations vanish in equilibrium and for $t=0$.

Substituting $\rho^{(k)}$ from Eq.~(\ref{eq14}) into
Eq.~(\ref{eq11}) and taking into account that the equilibrium
$k$-rod rotational current density vanishes we obtain the following
equation

\begin{equation} \label{eq15}
\frac{\partial} {\partial t} \, \rho^{(1)}_{n_1} \, (\Omega_1; t)
+ \nabla_1 \cdot \mathbf{j}_{n_1}^{(1)} (\Omega_1; t) =0
\end{equation}
where the 1-rod rotational current density reads

\begin{eqnarray} \label{eq16}
\mathbf{j}^{(1)}_{n_1} (\Omega_1; t) = - D_0 \left\{ \nabla_1
\rho_{n_1}^{ (1)} (\Omega_1; t) - \right. \nonumber \\ \left.
\sum\limits_{n_2 \neq n_1} \int
d \Omega _2 \mathbf{T}_{n_1 n_2} (\Omega_1, \Omega_2) \delta \rho^
{(2)}_{n_1 n_2} (\Omega_1, \Omega_2; t) \right\} .
\end{eqnarray}
At the next level, $k=2$, it follows by use of Eq.~(\ref{eq9})

\begin{eqnarray} \label{eq17}
\frac{\partial} {\partial t} \delta \rho^{(2)}_{n_1 n_2}
(\Omega_1, \Omega_2; t) = -\frac{1} {4 \pi} \, \mathbf{T}_{n_1,n_2}
(\Omega_1, \Omega_2) g^{(2)}_{n_1n_2} (\Omega_1, \Omega_2) \cdot
\mathbf{j}^{(1)}_{n_1} (\Omega_1; t) + \nonumber\\
+ D_0 \{\nabla_1 \cdot [\nabla_1 - \mathbf{T}_{n_1 n_2} (\Omega_1,
\Omega_2)] + (1 \leftrightarrow 2) \} \delta \rho ^{(2)}_{n_1 n_2}
(\Omega_1, \Omega_2; t) + \nonumber\\
+ A^{(2)}_{n_1 n_2} (\Omega_1, \Omega_2; t)
\end{eqnarray}
where

\begin{eqnarray} \label{eq18}
A^{(2)}_{n_1n_2} (\Omega_1, \Omega_2; t) = \nonumber \\
D_0 \nabla_1 \cdot \left[g
^{(2)}_{n_1 n_2} (\Omega_1, \Omega_2) \sum\limits_{n_3} \,
\frac{1} {4 \pi} \int d \Omega_3 \mathbf{T}_{n_1 n_3} (\Omega_1,
\Omega_3) \delta \rho^{(2)}_ {n_1 n_3} (\Omega_1, \Omega_3; t)\right]
\nonumber\\
-D_0 \sum\limits_{\nu=1}^{2} \, \nabla_\nu \cdot \sum\limits_{n_3}
\int d \Omega_3 \, \mathbf{T}_{n_\nu n_3} (\Omega_\nu, \Omega_3)
\delta \rho^{ (3)}_{n_1 n_2n_3} (\Omega_1, \Omega_2, \Omega_3; t)
\, .
\end{eqnarray}

Let us interpret these equations: Eq.~(\ref{eq15}) is the
continuity equation. The corresponding 1-rod current density
(Eq.~(\ref{eq16})) consists of two parts. The first describes the
contribution from the free Brownian dynamics of a tagged rod at
site $n_1$ and the second is due to the interaction of the tagged
rod with a second rod at side $n_2$. The first line on the r.h.s.
of Eq.~(\ref{eq17}) originates from the last term on the r.h.s. of
Eq.~(\ref{eq14}), its second line describes Brownian dynamics of
an \textit{isolated} pair of rods at sites $n_1$ and $n_2$, and the
last line contains the influence of a third rod at $n_3$ on the
rods at $n_1$ and $n_2$. In order to close this set of equations
we follow the strategy of refs. \cite{27,28} and approximate the
influence of a third rod by introducing an effective, nonlocal (in
angular space and time) diffusion tensor
$\mathbf{D}^{\mathrm{eff}}(\Omega, \Omega';t) $ and simultaneously
neglecting $A_{n_1n_2}^{(2)}$:

\begin{eqnarray} \label{eq19}
\left\{D_0 \nabla_1 \cdot[\nabla_1 - \mathbf{T}_{n_1 n_2} (\Omega_1,
\Omega_2)] + (1 \leftrightarrow 2) \right\} \delta \rho^{(2)}_{n_1 n_2}
(\Omega_1, \Omega_2; t) \rightarrow \nonumber\\
\left\{\nabla_1 \cdot \int\limits_0^t d t' \int d \Omega'_1
\mathbf{D}^{\mathrm{eff}} (\Omega_1, \Omega'_1; t-t') \cdot
[\nabla'_1 - \mathbf{T}_{n_1 n_2} (\Omega'_1, \Omega_2) ] \delta
\rho^{(2)}_{n_1 n_2} (\Omega'_1, \Omega_2;t) + \right. \nonumber\\
\left. + \nabla_2 \cdot \int\limits_0^t d t'\int d \Omega'_2
\mathbf{D}^{\mathrm{eff}}(\Omega_2, \Omega'_2; t-t') \cdot
[\nabla'_2 - \mathbf{T}_{n_2n_1} (\Omega'_2, \Omega_1)] \delta
\rho^{(2)}_{n_1 n_2} (\Omega_1, \Omega'_2; t) \right\}\nonumber\\
=:(\Lambda^{(2)} * \delta \rho^{(2)})_{n_1 n_2} (\Omega_1, \Omega_2;
t),
\end{eqnarray}

\begin{equation} \label{eq20}
A^{(2)}_{n_1n_2} (\Omega_1, \Omega_2; t) \approx 0.
\end{equation}

The generalized rotational diffusion tensor $\mathbf{D} (\Omega,
\Omega'; t)$ is defined through a constitutive equation relating
the 1-rod current density and the (angular) gradient of the 1-rod
density

\begin{equation} \label {eq21}
\mathbf{j}_{n_1}^{(1)} (\Omega; t) = - \int\limits_0^t d t' \int d
\Omega' \mathbf{D} (\Omega, \Omega'; t-t') \cdot \nabla_{\Omega'}
\rho_{n_1}^{(1)} (\Omega'; t).
\end{equation}

As the final approximation we impose a self-consistency condition,
\begin{equation} \label{eq22}
\mathbf{D}^{\mathrm{eff}} = \mathbf{D}.
\end{equation}

Eqs. (\ref{eq15}--\ref{eq17}), (\ref{eq19}--\ref{eq22})
form a closed set of equations for $\rho^{(1)}$, $\delta
\rho^{(2)}$ and $\mathbf{D}$. Taking their Laplace transform
\footnote{Here we use the definition $\widehat{f}(z)=
\int\limits^\infty_0 d t f(t) \exp (- z t)$, Re $z > 0$ and take
$\delta \rho^{(2)}_{n_1 n_2} (\Omega_1, \Omega_2; t=0)=0$ into
account.} it is straightforward to (formally) eliminate
$\rho^{(1)}$ and $\delta \rho^{(2)}$ which results in:

\begin{eqnarray} \label{eq23}
D_0^{-1} {\mathbf{\widehat{j}}}_{n_1} ( \Omega_1; z) \nonumber \\
+ \sum\limits_{n_3} \frac{1} {4 \pi} \int d \Omega_3
\mathbf{T}_{n_1 n_3} (\Omega_1, \Omega_3) [(z-
\widehat{\Lambda}^{(2)})^{-1} * (\mathbf{T}_{n_1 n_3} \,
g^{(2)}_{n_1 n_3} \cdot {\widehat{\mathbf{j}}}^{(1)}_{n_1})]_{n_1
n_3} (\Omega_1, \Omega_3; z)
\nonumber\\
= \int d \Omega_3 \,\, {\mathbf{\widehat{D}}}^{ -1} (\Omega_1,
\Omega_3; z) \cdot {\mathbf{\widehat{j}}}_{n_1} ( \Omega_3; z)
\end{eqnarray}
where \ $\widehat{}\;$  denotes the Laplace transformed
quantities. From Eq.~(\ref{eq23}) we find immediately the
self-consistent equation for ${\mathbf{\widehat{D}}} (\Omega_1,
\Omega_2;z)$:

\begin{eqnarray} \label{eq24}
{\mathbf{\widehat{D}}}^{-1} (\Omega_1,
\Omega_2;z)=\mathbf{D}^{-1}_0 \delta
(\Omega_1|\Omega_2) \nonumber \\
+ \frac{1} {4 \pi} \, \sum\limits_{n_3} \, \mathbf{T}_{n_1 n_3}
(\Omega_1, \Omega_2) [ (z- \widehat{\Lambda}^{(2)})^{-1} *
(\mathbf{T}^t_{n_1n_3} \, g^{(2)}_{n_1 n_3} )]_{n_1n_3} (\Omega_1,
\Omega_2;z)
\end{eqnarray}

Taking the limit $z \rightarrow 0$ in Eq.~(\ref{eq24}) and
operating with $(1/4 \pi) \int d \Omega_1 \int d \Omega_2$ on both
sides of the resulting equation yields the following 
for the diffusion tensor $\mathbf{D}
(\Omega_1, \Omega_2) \equiv {\mathbf{\widehat{D}}} (\Omega_1,
\Omega_2; z=0)$:

\begin{equation} \label{eq25}
4 \pi \langle \mathbf{D}^{ -1} \rangle = \mathbf{D}_0^{-1} -
\sum\limits_{n_3} \langle ({\widehat{\Lambda}}^{(2)\dagger})^{-1}*
\mathbf{T})_{n_1 n_3} \, \mathbf{T}^t_{n_1 n_3} \rangle
\end{equation}
with

\begin{eqnarray} \label{eq26}
\langle f_{n_1 n_2} \, h_{n_1 n_2} \rangle \nonumber \\
= \frac{1} {(4 \pi)^2}
\, \int d \Omega_1 \int d \Omega_2  g^{ (2)}_{n_1 n_2} (\Omega_1,
\Omega_2)
 f^*_{n_1 n_2}(\Omega_1, \Omega_2)  h_{n_1
n_2} (\Omega_1, \Omega_2),
\end{eqnarray}
and

\begin{equation} \label{eq27}
(\mathbf{D}_0)^{\alpha \beta} \quad \quad = D_0 \, \delta ^{\alpha
\beta}.
\end{equation}
$\mathbf{T}^t$ is the transpose of $\mathbf{T}$ and for the
adjoint 2-rod operator we find from Eq.~(\ref{eq19}):

\begin{eqnarray} \label{eq28}
({\widehat{\Lambda}}^{ (2) \dagger} * f)_{n_1 n_2}
(\Omega_1,\Omega_2)
\nonumber \\
= [\nabla_1 + \mathbf{T}_{n_1 n_2} (\Omega_1, \Omega_2)]
\cdot \int  d\Omega_3 \mathbf{D} (\Omega_3, \Omega_1) \cdot
\nabla_3
f_{n_1 n_2} (\Omega_3, \Omega_2) \nonumber\\
+ [\nabla_2 + \mathbf{T}_{n_2 n_1} (\Omega_2, \Omega_1)] \cdot
\int d \Omega_3 \mathbf{D} (\Omega_3, \Omega_2) \cdot \nabla_3
f_{n_1 n_2} (\Omega_1, \Omega_3).
\end{eqnarray}
The reader should note that (i) $g^{(2)}_{n_1 n_2} (\Omega_1,
\Omega_2)$ in Eq.~(\ref{eq26}) can be skipped, because it is equal
to one almost everywhere; of course, it must not be dropped if
$\nabla_1$ or $\nabla_2$ act on it, like in Eq.~(\ref{eq18}); in
that case one can use Eq.~(\ref{eq9}); (ii) $\langle
\mathbf{D}^{-1} \rangle$ does not depend on $n_1$ and $n_2$ and
(iii) the tensor $\langle (( {\widehat{\Lambda}}^{ (2)
\dagger})^{-1} * \mathbf{T})_{n_1 n_3} \, \mathbf{T}^t _{n_1 n_3}
\rangle$ in Eq.~(\ref{eq25}) depends only on
$\mathbf{R}_{n_3}-\mathbf{R}_{n_1}$ because of the lattice
translational invariance. Accordingly $\sum\limits_{n_3} \langle
\cdots \rangle$ does not depend on $n_1$. Eq.~(\ref{eq25}) can
also be rewritten as follows

\begin{equation} \label{eq29}
4 \pi \langle \mathbf{D} ^{-1} \rangle = \mathbf{D}_0^{-1} +
\sum\limits_{n_3} \int\limits_0^{\infty} dt \langle
(e^{{\widehat{\Lambda}}^{(2) \dagger} t} * \mathbf{T})_{n_1 n_3}
\, \mathbf{T}_{n_1 n_3}^t \rangle .
\end{equation}

Let us discuss Eq.~(\ref{eq29}). It is a \textit{functional
equation} for the rotational diffusion tensor $\mathbf{D}$, since
${\widehat{\Lambda}}^{(2) \dagger}$ also involves $\mathbf{D}$. It
$ l \leq 1$ then $\mathbf{T}_{n_1 n_2} (\Omega_1, \Omega_2) \equiv
0 $ and therefore Eq.~(\ref{eq29}) implies that

\begin{equation} \label{eq30}
\mathbf{D} (\Omega_1, \Omega_2) = \mathbf{D}_0 \delta (\Omega_1 |
\Omega_2)
\end{equation}
as it should be. Increasing $l$ beyond one leads to an
``increase'' of the friction tensor $\langle
\mathbf{D}^{-1}\rangle$, due to the positive definite,
time-dependent torque-torque correlation tensor $\langle (
\exp({\widehat{\Lambda}}^{(2) \dagger} t) * \mathbf{T})_{n_1 n_3}
\, \, \mathbf{T}^t _ {n_1 n_3} \rangle$ which determines the
second term in Eq.~(\ref{eq29}). This ``increase'' of the
``renormalized'' friction tensor $\mathbf{D}^{-1}$ implies a
decrease of the ``renormalized'' diffusion tensor $\mathbf{D}$.
Since this one enters into the exponent of $\exp (
{\widehat{\Lambda}}^{ (2) \dagger} t)$ the relaxation of the
torque-torque correlation will slow down. This in turn leads to an
``increase'' of the second term of Eq.~(\ref{eq29}) leading to a
further ``increase'' of $\mathbf{D}^{-1}$ and so on. This
feedback mechanism, which is different from but still resembles that
of MCT \cite{13,14,15}, finally may lead to the vanishing of
diffusion and therefore to a glass transition.

Although the functional equation possesses a rather clear
structure it probably cannot be solved exactly. For this one would have
to determine the eigenvalues and eigenfunctions of $
{\widehat{\Lambda}}^{ (2) \dagger}$. Due to the singular character
of $\mathbf{T}_{n_1n_2} (\Omega_1, \Omega_2)$ (cf.
Eq.~(\ref{eq8})) on the three-dimensional contact-hypersurface
defined by Eq.~(\ref{eq1}) this does not seem to be feasible. In
order to make progress and to get explicit results for
$\mathbf{D}$ we will use the following additional approximation:

\begin{equation} \label{eq31}
(\mathbf{D} (\Omega_1, \Omega_2)) ^{\alpha \beta} \approx D(l) \delta
^{\alpha \beta} \delta (\Omega_1 | \Omega_2).
 \end{equation}
Substituting Eq.~(\ref{eq31}) into Eq.~(\ref{eq29}) and acting
with $\frac{1}{3} \sum\limits_{\alpha, \beta}$ on both sides
yields:

\begin{equation} \label{eq32}
D(l) =D_0 [1 - \upsilon(l)] \quad \quad .
\end{equation}
The l-dependent coupling function $\upsilon(l)$
follows from the second term of
Eq.~(\ref{eq29}) by use of Eqs.~(\ref{eq28}) and (\ref{eq31}):

\begin{equation} \label{eq33}
\upsilon(l)= \frac{1} {3} \sum\limits_{n_3} \,
\int\limits_0^{\infty} d t \langle \mathbf{T}_{n_1 n_3} \cdot
e^{\mathcal{L}^{(2) \dagger} t} \, \mathbf{T}_{n_1 n_3} \rangle
\end{equation}
with the adjoint of the 2-rod Smoluchowski operator (cf.
Eq.~(\ref{eq12})):

\begin{equation} \label{eq34}
\mathcal{L}^{(2) \dagger} = [ \nabla_1 + \mathbf{T}_{n_1 n_2}
(\Omega_1, \Omega_2)] \cdot \nabla_1 + (1 \leftrightarrow 2)
\end{equation}
where $D_0$ is replaced by one. The calculation of the relevant
quantity $\upsilon(l)$ and the determination of a  critical length
$l_c$ at which $D(l)$ vanishes will be presented in the next
section.

\section{RESULTS}
%\smallskip
%{\bf\textrm{A. Glass transition singularity}}\\
%\smallskip
\subsection{Glass transition singularity}

Using the local approximation Eq.~(\ref{eq31}), we obtained
Eq.~(\ref{eq32}) which constitutes a rather simple result. Before
we calculate $\upsilon(l)$ let us discuss $D(l)$ and $\upsilon(l)$
on a qualitative level. Since $\mathbf{T}_{n_1 n_2} (\Omega_1,
\Omega_2)\equiv 0$ for $l \leq 1$ it follows that $\upsilon(l)=0$
and therefore $D(l)=D_0$ for $l \leq 1$, as it should be. Next, we
introduce:

\begin{equation} \label{eq35}
\upsilon_0 (l/r_n) = \frac{1}{3} \, \int\limits_0^{\infty} \, d t
\langle \mathbf{T}_{\mathrm{0n}} \cdot e^{\mathcal{L}^{(2)
\dagger}t} \, \mathbf{T}_{\mathrm{0n}} \rangle
\end{equation}
where $r_n= |\mathbf{R}_n| /a$. This is related to the
torque-torque correlator of one rod at the origin and another one
at site $n$. Note that the r.h.s. of Eq.~(\ref{eq35}) depends on
$|\mathbf{R}_n|$, but not on the direction of $\mathbf{R}_n$.
$\upsilon_0 (x)$, of course, vanishes for $x \leq 1$, is positive
for $x > 1$ (see below) and we expect it to converge to a finite limit
$\upsilon_0 ^{\infty}$ for $x \rightarrow \infty$. $\upsilon (l)$
is completely determined by $\upsilon_0(x)$:

\begin{equation} \label{eq36}
\upsilon(l)= \sum\limits_n \,\, \upsilon_0 (l /r_n)
\end{equation}
where the sum is restricted to such $n$ for which $r_n < l$. This
relation can be used to determine the asymptotic behavior of
$\upsilon(l)$ for $l \rightarrow \infty$. Approximating the sum by
an integral which becomes more and more accurate with increasing
$r_n$ we get:

\begin{eqnarray*}
 \upsilon(l) \approx \int\limits_{1 \leq r \leq l} d^3 r \,
\upsilon_0 (l/r) = 4 \pi \int\limits_1 ^{l} d r \,\, r^2 \,
\upsilon_0 (l/r)
\nonumber\\
= 4 \pi \left(\;\;\int\limits_{l^{-1}}^1 \, dx x^2 \, \upsilon_0
\left(\frac{1} {x}\right) \right) l^3
\end{eqnarray*}
which gives

\begin{equation} \label{eq37}
\upsilon(l) \approx c \cdot l^3 + O(l^2) \quad , \quad l
\rightarrow \infty
\end{equation}
with

\begin{equation} \label{eq38}
c= 4 \pi \int\limits_0^1 dx x^2 \, \,  \upsilon_0
\left(\frac{1} {x}\right) > 0 \, \, .
\end{equation}

That $\upsilon(l) \sim l^3$ for $l \rightarrow \infty$ is obvious
since all rods within a sphere of radius of order $l$ will collide
with the central one. Their number, of course, is proportional to
$l^3$. Because $\upsilon(l)=0$ for $l \leq 1$ and $\upsilon(l)$
increases linearly with $l^3$ for large $l$ there must exist a
critical length $l_c$ for which $\upsilon(l_c)=1$ and therefore
$D(l_c)=0$. To determine $l_c$ at which a dynamical glass
transition takes place we have to calculate $\upsilon_0(x)$.
Although this quantity is much simpler than the second term in
Eq.~(\ref{eq29}), it cannot be determined exactly analytically (see
below). Therefore the exact evaluation can only be done
numerically. For such a numerically exact calculation we performed
a simulation of the Brownian dynamics (defined by
$\mathcal{L}^{(2) \dagger})$ of  an \textit{isolated} pair of rods
with distance $r_n$. The time-dependent torque-torque correlator
resulting from this simulation is shown in Fig.~\ref{fig5} for
several values of $l/r_n$. For $t \rightarrow 0$ we find a power
law divergence $t^{-1/2}$ as expected and already proven for the
force-force correlator of hard spheres \cite{37,38}. Since only
two rods are considered, the system is ergodic and therefore the
torque-correlations relax to zero. Figure ~\ref{fig5} shows that
the relaxational behavior changes qualitatively around $l/r_n=2$
from a fast decay like for free Brownian dynamics to a rather slow
relaxation. This crossover at $l/r_n \approx2$ is related to the
properties discussed in section 2. Whether the \textit{long time}
decay of the torque-torque correlation is purely exponential,
proportional to $t^{-\alpha}\,  \exp (- \lambda t)$ or even a
single power law $t^{-\alpha}$ (as it is true for two hard spheres
\cite{37}) cannot be decided from our numerical data. Although the
log-linear representation in Figure 5 exhibits a bending for
large $t$ (which would indicate deviation from a pure exponential)
the exponential behavior could appear on a much larger time scale
on which the statistical fluctuations of the numerical data
prevent the determination of the precise long time decay. The
numerical evaluation of the time integral in Eq.~(\ref{eq35})
leads to $\upsilon^{\mathrm{num}}_0 (x)$ presented in Figure
~\ref{fig6}. The reader should note the strong increase starting
at $x=l/r_n \approx 2$.

Now let us turn to the analytical calculation of $\upsilon_0(x)$.
For this we rewrite $\upsilon_0(x)$ as

\begin{equation} \label{eq39}
\upsilon_0(x) = \frac{1} {3} \langle \mathbf{T}_{\mathrm{0n}}
\cdot (- \mathcal{L}^{(2) \dagger})^{ -1} \mathbf{T}_{\mathrm{0n}}
\rangle.
\end{equation}
which is (up to the factor $1/3$) 
the Laplace transformed torque-torque correlator at $z=0$.

Analogous to refs. \cite{27,28} we introduce a vector function
$\mathbf{f}_{n_1 n_2} (\Omega_1, \Omega_2)$ such that:

\begin{equation} \label{eq40}
\mathcal{L}^{ (2)\dagger} \mathbf{f}_{n_1 n_2} (\Omega_1,
\Omega_2) = \mathbf{T}_{n_1 n_2} (\Omega_1, \Omega_2).
\end{equation}
Eq. (\ref{eq40}) can be decomposed into a \textit{regular} part

\begin{equation} \label{eq41}
[\nabla_1^2 + \nabla_2^2] \mathbf{f}_{n_1 n_2} (\Omega_1,
\Omega_2)=0,
\end{equation}
and a singular one:

\begin{equation} \label{eq42}
[s_{12}(\mathbf{u}_1 \times {\mathbf{\widehat{r}}}_{12}^{\bot} )
\cdot \mathbf{\nabla}_1 + s_{21} (\mathbf{u}_2 \times
{\widehat{\mathbf{r}}}_{21}^{\bot}) \cdot \mathbf{\nabla}_2]\,\,\,
\mathbf{f}_{n_1 n_2} (\Omega_1, \Omega_2)=s_{12} (\mathbf{u}_1
\times {\mathbf{\widehat{r}}}_{12} ^{\bot}).
\end{equation}
Eq.~(\ref{eq42}) is to be satisfied for $\Omega_1, \Omega_2$
located on the three-dimensional contact hypersurface.
Accordingly, Eqs.~(\ref{eq41}-\ref{eq42}) describe a
boundary value problem with boundary value on the
three-dimensional hypersurface embedded in the four-dimensional
space built by the surface of two unit spheres. This is a
difficult mathematical problem which seems to resist a rigorous
solution. If it could be solved then the calculation of
$\upsilon_0(x)$ is reduced to the calculation of an integral over
$\Omega_1$ and $\Omega_2$:

\begin{equation} \label{eq43}
\upsilon_0(x) = -\frac{1} {3} \langle \mathbf{T}_{\mathrm{0n}}
\cdot \mathbf{f}_{\mathrm{0n}} \rangle.
\end{equation}
Therefore, we resort to an alternative. It is easy to prove (by the
use of Eq.~(\ref{eq9})) that Eq.~(\ref{eq40}) is the variational
equation of the functional:

\begin{equation} \label{eq44}
\mathcal{F} [ \mathbf{f}_{n_1 n_2} (\Omega_1, \Omega_2) ] =
\frac{1} {3} [ \langle \mathbf{f}_{n_1 n_2} \cdot \mathcal{L}^
{(2) \dagger} \mathbf{f}_{n_1 n_2} \rangle - 2 \langle
\mathbf{T}_{n_1 n_2} \cdot \mathbf{f}_{n_1 n_2} \rangle ] \quad.
\end{equation}
For a similar discussion see ref. \cite{40}. It can be easily shown
by partial integration that:

\begin{eqnarray} \label{eq45}
\langle \mathbf{f}_{n_1 n_2} \cdot \mathcal{L}^{(2) \dagger}
\mathbf{f}_{n_1 n_2} \rangle= \nonumber\\
-\frac{1}{(4\pi)^2}
\int d \Omega_1 \int d \Omega_2 g^{(2)}_{n_1 n_2} (\Omega_1,
\Omega_2) \sum\limits_{\alpha} \left[ ( \nabla_1 f^{\alpha}_{n_1n_2}
(\Omega_1, \Omega_2))^2 \right. \nonumber\\
\left. + (\nabla_2 f^{\alpha}_{n_1
n_2} (\Omega_1, \Omega_2))^2 \right]
\end{eqnarray}
which is negative for all $\mathbf{f}_{n_1n_2} \neq
\mathrm{const}$, \textit{i.e.} $\mathcal{L}^{(2) \dagger}$ is a negative
operator. Therefore one can follow ref. \cite{40} and prove that
the solution $\mathbf{f}_{n_1 n_2}^{\mathrm{sol}} (\Omega_1,
\Omega_2)$ of Eq.~(\ref{eq40}) is the non-degenerate maximum of
$\mathcal{F}$. From Eqs.~(\ref{eq40}), ~(\ref{eq43}) and
~(\ref{eq44}) we then find that

\begin{equation} \label{eq46}
\upsilon_0(x) = \mathcal{F} [ \mathbf{f}_{n_1
n_2}^{\mathrm{sol}} (\Omega_1, \Omega_2)],
 \end{equation}
\textit{i.e.} the maximum value of $\mathcal{F}$ is just $\upsilon_0(x)$.
Since the maximum is non-degenerate the following inequality is true
for any trial function $\mathbf{f}_{n_1 n_2}^{\mathrm{var}}
(\Omega_1, \Omega_2)$

\begin{equation} \label{e47}
\upsilon_0^{\mathrm{var}} (x) \equiv \mathcal{F} [ 
\mathbf{f}_{n_1 n_2}^{\mathrm{var}} (\Omega_1, \Omega_2) ] \leq
\upsilon_0 (x).
\end{equation}
Choosing an appropriate trial function offers the possibility to
determine a lower bound for $\upsilon_0(x)$ and therefore an upper
bound for the critical length $l_c$.

The boundary value problem described above is similar to those
arising from electrodynamics or hydrodynamics. Its singular part
implies that $\mathbf{f}_{n_1 n_2} (\Omega_1, \Omega_2)$ is
discontinuous at the contact surface. This is similar to the
behavior of the electric field at a dipolar layer. Without
restricting generality, we can choose $\mathbf{R}_{n_2}-
\mathbf{R}_{n_1}$ along the $z$-axis. Then the relevant coordinate
is $\phi=\phi_2-\phi_1$ and contacts can occur at $\phi=0^{\pm}$
or $\phi= \pi ^{\pm}$ (see discussion in section 2), \emph{i.e.}
$\mathbf{f}_{n_1 n_2} (\Omega_{n_1}, \Omega_{n_2}) \equiv
\mathbf{f}_{n_1 n_2} (\theta_1, \theta_2, \phi_1, \phi)$ must be
discontinuous at $\phi= 0$ and $\phi= \pi$. Next, Eq.~(\ref{eq43})
makes it obvious that $|\mathbf{T}_{n_1 n_2} \cdot \mathbf{f}_{n_1
n_2}|$ should be made as large as possible. This can be done by
choosing $\mathbf{f}_{n_1 n_2} \sim \mathbf{T}_{n_1 n_2} \sim
\mathbf{u}_1 \times {\mathbf{\widehat{r}}}_{12} ^{\bot} = (\cos
\theta_1 \, \cos \phi_1, \, \cos \theta_1 \, \sin \phi_1,
-\sin\theta_1).$ Finally, the calculation of $\mathcal{F}$ from
Eq.~(\ref{eq44}) requires the calculation of the r.h.s. of
Eq.~(\ref{eq45}). Since $\nabla_i f^{\alpha} _{n_1 n_2} (\Omega_1,
\Omega_2)$ involves $1/ \sin \theta_i$ for $x-$ and $y-$
component of the gradient, the
integrals in Eq.~(\ref{eq45}) will not exist unless 
$f^{\alpha}_{n_1 n_2} \sim (\sin \theta_1)^{\nu_1} (\sin
\theta_2)^{\nu_2}$ with $\nu_1 \geq 1$ and $\nu_2 \geq 1$. These
considerations suggest to use the following trial function:

\begin{eqnarray} \label{eq48}
\mathbf{f}_{n_1 n_2}^{\mathrm{var}}(\Omega_1, \Omega_2)
&\equiv& \lambda \mathbf{h}_{n_1 n_2}^{\mathrm{var}}(\Omega_1, \Omega_2) 
=\lambda
\sin\theta_1 \sin \theta_2 \left (
\begin{array} {*{3}c@{\;}} \cos \theta_1  \cos \phi_1 \\ \cos \theta_1
\sin \phi_1 \\ -\sin\theta_1
\end{array} \right ) \nonumber \\ &&\times \left \{
\begin{array}{ll} \frac{\pi}{2} -(\phi_2-\phi_1),  & 0 < \phi_2-\phi_1 <
\pi
\\
\frac{3 \pi}{2} - (\phi_2- \phi_1), & \pi < \phi_2-\phi_1 < 2 \pi
\end{array} \right.
\end{eqnarray}
with $\lambda$ being a variational parameter. For the
$(\phi_2-\phi_1)$-dependence we have made the simplest choice
taking a linear variation with discontinuities at $0$ and $\pi$.
Note also that 
$\mathbf{f}_{n_1 n_2}^{\mathrm{var}} (\Omega_1,\Omega_2)$ does not
depend on either $\mathbf{R}_{n_2} - \mathbf{R}_{n_1}$ or $x=l/r_n$,
which will be not true for the true, exact solution
$\mathbf{f}^\mathrm{sol}_{n_1 n_2}
(\Omega_1, \Omega_2)$. Substituting $ \mathbf{f}^\mathrm{var}_{n_1
n_2} (\Omega_1, \Omega_2)$ into Eq.~(\ref{eq44}), where the first
term is evaluated by Eq.~(\ref{eq45}), one obtains:

\begin{equation} \label{eq49}
\mathcal{F} [\mathbf{f}^{\mathrm{var}}_{n_1 n_2} (\Omega_1,
\Omega_2)  ] = F( \lambda) = \lambda^2 I_2 - 2 \lambda I_1 (x)
\end{equation}
with

\begin{equation} \label{eq50}
I_1 (x)= \frac{1} {3} \frac{1} {(4 \pi)^2} \int d \Omega_1 \int d
\Omega_2 \,\, \mathbf{h}^{\mathrm{var}}_{n_1 n_2} (\Omega_1,
\Omega_2) \cdot \mathbf{T}_{n_1 n_2} (\Omega_1, \Omega_2)
\end{equation}

\begin{eqnarray} \label{eq51}
I_2 = - \frac{1} {3} \frac{1} {(4 \pi)^2} \nonumber \\ \times
\int d \Omega_1 \int d
\Omega_2 \sum\limits_{\alpha} \left[(\nabla_1 h^{{\mathrm{var}},
\alpha}_{n_1 n_2} (\Omega_1, \Omega_2))^2 + (\nabla_2
h^{{\mathrm{var}}, \alpha}_{n_1 n_2} (\Omega_1, \Omega_2))^2\right].
\end{eqnarray}
Note that $I_2$ does not depend on $x$. $F(\lambda)$ is easily
maximized. As a result one obtains:

\begin{equation} \label{eq52}
\upsilon_0^{\mathrm{var}} (x)= F (\lambda_{\mathrm{max}}) =
\frac{(I_1(x))^2} {(-I_2)} \quad .
\end{equation}
$I_2$ can be calculated analytically:

\begin{equation} \label{eq53}
I_2= - \frac{2} {81} (27 + 2 \pi^2) \cong -1.154
\end{equation}
whereas $I_1(x)$ must be calculated by numerical integration, due
to the nontrivial contact hypersurface. The result
$\upsilon_0^{\mathrm{var}} (x)$ following from this approach is
shown in Fig.~\ref{fig6}. It is interesting that
$\upsilon_0^{\mathrm{var}} (x)$ reproduces the crossover behavior
found for $\upsilon_0^{\mathrm{num}} (x)$ close to $x=2$ from
$\upsilon_0^{\mathrm{var}}(x) \approx 0$ for $x<2$ to $\upsilon_0
^{\mathrm{var}} (x) > 0$ for $x > 2$. Comparison of
$\upsilon_0^{\mathrm{num}} (x)$ with $\upsilon_0^{\mathrm{var}}
(x)$ confirms that $\upsilon_0^{\mathrm{var}} (x)$ is a lower
bound for $\upsilon_0^{\mathrm{num}}(x)$, as it should be.
Introducing $\upsilon_0 (x)$ from both approaches into
Eq.~(\ref{eq36}) leads to $\upsilon(l)$ represented in
Fig.~\ref{fig7} for an sc-lattice. From this figure we get:

$$l^{\mathrm{num}}_c \cong 3.45 \quad , \quad l_c^{\mathrm{var}}
\cong5.68 \quad : \quad \mathrm{sc\ lattice}. $$
$\upsilon(l)$ can be calculated for any periodic lattice. For the
other cubic lattices we obtained

\begin{eqnarray*}
l_c^{\mathrm{num}} \cong \left \{
\begin{array}{ll} 2.78 & \\ 2.20 & \end{array}  \right.
, \quad  l_c^{\mathrm{var}} \cong
 \left \{
\begin{array}{ll} 4.84 & \mathrm{bcc\ lattice}\\
3.96 & \mathrm{fcc\ lattice} .
\end{array} \right.
\end{eqnarray*}

Since $\upsilon^ {\mathrm{var}} (l) \leq \upsilon^{\mathrm{num}}
(l) \cong \upsilon (l)$, it follows that $l_c^{\mathrm{var}} \geq
l^{\mathrm{num}}_c \cong l_c$. The critical length decreases when
going from sc to bcc to fcc lattice. This decrease is related
to (i) the increase of the coordination number $z$ from 6 to 8 to
12 and (ii) a decrease of the nearest neighbor distances $r_1$
from 1 to $\sqrt{3}/2$ to $\sqrt{2}/2$ (in units of the lattice
constant $a$). Since an increase of $z$ and a decrease of $r_1$
results in an increase of the steric hindrance, $l_c$ must
decrease. As the increase of steric hindrance is equivalent to an
increase of collisions, \textit{i. e.} of contacts, our result has some
similarity to that found in ref. \cite{41}. These authors have
used a probabilistic approach in order to derive a criterion for
the \textit{mechanical} stability of an off-lattice system of
infinitely thin hard rods in its randomly closed-packed glassy
state. It has been found that a mechanically stable amorphous
phase occurs if the average contact point per rod becomes 5.
However, such an arrangement of hard rods with $d=0$ is
dynamically unstable, as discussed in the Introduction. In the
simulation performed in ref. \cite{30} it was found
$l_c^{\mathrm{sim}} \cong 2.7$ for an fcc lattice. On the other
hand ref. \cite{31} found $l_c^{\mathrm{sim}} \cong 4.5$
for an sc lattice. Note that in ref. \cite{31} the rods were
fixed with one of their endpoints, and not with their centers.
Our result for $l_c^{\mathrm{num}}$ and
$l_c^\mathrm{var}$ agree satisfactorily to the simulational
result of ref. \cite{30}. The difference between our result and
that of ref. \cite{31} may be due to the different way of fixing
rods on the lattice that was mentioned above.

The authors of ref. \cite{30} have also studied the $l-$dependence
of the rotational diffusion constant $D(l)$ defined by:

\begin{equation}  \label{eq54}
D(l) = \lim\limits_{t \rightarrow \infty} \left( - \frac{1} {2 t}
\ln \left[\frac{1} {N} \sum\limits_{n=1}^{N} \langle \mathbf{u}_n
(t) \cdot \mathbf{u}_n (0) \rangle \right]\right) \quad.
\end{equation}
The variation of $D$ over about two orders in magnitude follows a
power law $D(l) \sim (l_c - l )^{\gamma^{MD}} $ with $\gamma^{MD}
\cong 4.2$. Since $\upsilon(l)$ is analytical for $l > 1$ it
follows that $1-\upsilon(l) \sim l_c -l$ and therefore $D(l) \sim
(l_c -l)$. Hence our analytical theory yields $\gamma=1$ which
differs significantly from $\gamma^{MD}$. This deviation may have
two reasons. First, our result $\gamma=1$ is mean-field-like (see,
\emph{e.g.}, ref. \cite{23}). Note that in ref. \cite{23} going
beyond mean field approximation it has been found $ \gamma \cong
7/6$ \cite{23} which is still close to one. Second, $\gamma^{MD}
\cong 4.2$ is unusually high. Simulations and experiments of
supercooled liquids usually yield $\gamma \approx 2$ consistent
with most MCT-analyses \cite{13,14,15, 17,18,19}.

\subsection{Nonergodicity parameter and dynamics close to $l_c$}

The analysis in subsection 4.1 has proven that our theory predicts
a dynamical
glass transition at a critical length $l_c$. There are two important
questions remaining. First, does the nonergodicity parameter 
change at $l_c$ in a continuous (type-A transition \cite{13}) 
or discontinuous (type-B transition \cite{13}) way? 
Second, what is the time-- or
frequency-dependence of the 1-rod density $\rho^{(1)}_n$ close to
$l_c$? The present subsection will give answers to both
questions.

Let us expand $\widehat{\rho}_n ^{(1)} (\Omega; z)$ and
${\bf{\widehat{j}}}_n^{(1)} (\Omega; z)$ into spherical harmonics:
\begin{equation} \label{eq55}
\widehat{\rho}^{(1)}_n (\Omega; z) = \sum\limits_\lambda \,\,
\widehat{\rho}^{(1)}_{n, \lambda} (z) \, Y_\lambda (\Omega)
\end{equation}

\begin{equation} \label{eq56}
{\bf\widehat{j}}^{(1)}_n (\Omega; z) = \sum\limits_\lambda \,\,
{\bf\widehat{j}}^{(1)}_{n,\lambda} (z) \, Y_\lambda (\Omega).
\end{equation}
where $\lambda=(j,m)$, $j=0,1,2, \cdots ; $ $-j \leq m \leq j$.
Furthermore, let us extend approximation (\ref{eq31}) to finite
$z$:

\begin{equation} \label{eq57}
({\bf\widehat{D}} (\Omega_1, \Omega_2; z))^{\alpha \beta} \approx
\widehat{D}(z) \delta^{\alpha \beta} \, \delta (\Omega_1 |
\Omega_2).
\end{equation}
It is easy to prove then 
that the Laplace transform of Eqs.~(\ref{eq15}) and
~(\ref{eq21}) leads to

\begin{equation} \label{eq58}
{\widehat\phi}_\lambda (z) \equiv \widehat{\rho}_\lambda^{(1)} (z)
/ \rho_\lambda^{(1)} (t=0)=1 / [z + j (j + 1) \widehat{D} (z)],
\end{equation}
where we skipped the index $n$.

The nonergodicity parameter $f_\lambda$ is defined as follows:
\begin{equation} \label{eq59}
f_\lambda = \lim\limits_{z \rightarrow 0} \, z
\widehat{\phi}_\lambda (z).
\end{equation}

Taking into account approximation Eq.~(\ref{eq57}) we get from
Eq.~(\ref{eq24}) after operating with (1/4 $\pi$) $\int d \Omega_1
\, \int d \Omega_2$ on both sides the self-consistency equation
for $\widehat{D}(z)$:
\begin{equation} \label{eq60}
\frac{1} {\widehat{D}(z)} = \frac{1} {D_0} + \frac{1}
{\widehat{D}(z)} \, \widehat{\upsilon} (\zeta;l), \quad \zeta=
z/\widehat{D}(z),
\end{equation}
with
\begin{equation} \label{eq61}
\widehat{\upsilon} (\zeta;l)= \frac{1} {3} \, \sum\limits_n \, \langle
\mathbf{T}_{\mathrm{0n}} \cdot [\zeta - \mathcal{L}^{(2) \dagger}
]^{-1} \, \mathbf{T}_{\mathrm{0n}} \rangle.
\end{equation}
Inverse Laplace transform of Eq.~(\ref{eq61}) yields
\begin{equation} \label{eq62}
\upsilon(t;l) = \frac{1} {3} \, \sum\limits_n \, \langle
\mathbf{T}_{\mathrm{0n}} \cdot e^{\mathcal{L}^{{(2)} \dagger} t}  \,
\mathbf{T}_{\mathrm{0n}} \rangle.
\end{equation}
Note that $\upsilon(t;l)$ is the time-dependent analogue
of $\upsilon(l)$ defined in Eq. (\ref{eq33}).
 
Eq.~(\ref{eq60}) has a similar mathematical structure as
the corresponding equation for a Lorentz gas obtained from a mode
coupling approximation \cite{41a,42}. Assuming that $F(t)$ does not
decay slower than $t^{-5/2}$ for  $t \rightarrow \infty$ \cite{43}
it follows from the Tauberian theorems \cite{44} that
\begin{equation} \label{eq63}
\widehat{\upsilon} (\zeta;l)= \widehat{\upsilon}(0;l) + 
\widehat{\upsilon}' (0;l) \zeta + 
O(\zeta^{3/2})
\end{equation}
where
\begin{eqnarray} \label{eq64}
&&\widehat{\upsilon}(0;l) \equiv \upsilon(l)\nonumber\\
&&\widehat{\upsilon}' (0;l) = - \frac{1} {3} \, \sum\limits_n \langle
\mathbf{T}_{\mathrm{0n}} \cdot (\mathcal{L}^{{(2)}\dagger})^{-2}
\mathbf{T}_{\mathrm{0n}}\rangle := - a(l) < 0 .
\end{eqnarray}

Substituting Eq.~(\ref{eq63}-\ref{eq64}) into ~(\ref{eq60}) and
neglecting $O(\zeta^{3/2})$ we get a quadratic equation for
$z/\widehat{D}(z)$, the physical solution of which is
\begin{equation} \label{eq65}
z/ \widehat{D} (z) = \frac{1} {2} [ \epsilon(l) +
\sqrt{\epsilon^2 (l) + z \cdot t_0}]
\end{equation}
with
\begin{eqnarray} \label{eq66}
\epsilon(l)= \frac{\upsilon(l)-1} {a (l_c)} = \left
\{\begin{array}{ll}  < 0, \quad   l <l_c \quad  \textrm{(ergodic 
phase)} \\
\geq 0, \quad  l \geq l_c \quad  (\textrm{glass phase})\\
\end{array} \right.
\end{eqnarray}
and $t_0=4/ (a (l_c) D_0)$ being a microscopic time scale. Note that
Eq.~(\ref{eq65}) is identical to the corresponding equation
obtained for the so-called $F_1$-model \cite{13}. Accordingly, all
results for the $F_1$-model \cite{13} hold here as well. This
implies (i) that the  glass transition is of type A, \textit{i.e.} 
the nonergodicity parameters (obtained from Eqs.~(\ref{eq58}) and
~(\ref{eq59}))
\begin{eqnarray} \label{eq67}
f_\lambda(l)= \left \{\begin{array}{ll} 0 \quad  \quad \quad \quad
\quad 
\quad \quad \quad \quad \quad \quad \quad \quad \, , \quad l =
\leq l_c\\
\displaystyle
\frac{\epsilon (l)}{j(j+1)} \cong
\frac{\upsilon'(l_c)}{a(l_c)
j(j +1)} (l-l_c) , \quad l \rightarrow l_c^+\\
\end{array} \right.
\end{eqnarray}
vary continuously at $l_c$, (ii) at the critical point $l=l_c$
and for $z \rightarrow 0$, $t \rightarrow \infty$
\begin{equation}  \label{eq68}
\widehat{\phi}_\lambda (z)\cong\frac{1} {2} (z t_0)^{1/2},
\quad \phi_\lambda (t) \cong \frac{1}{\sqrt{4 \pi}} (t/t_0)^{-1/2}
\end{equation}
and (iii) for $|l-l_c| \ll 1$ and for $z \rightarrow 0$, $t
\rightarrow \infty$
\begin{equation} \label{eq69}
\widehat{\phi}_\lambda (z, \epsilon) = (|\epsilon| /
\omega_\epsilon) \, \widehat{\phi}_\lambda (z/
\omega_\epsilon), \quad \phi_\lambda (t, \epsilon) =
|\epsilon| \phi_\lambda (t/t_\epsilon)
\end{equation}
with $\omega_\epsilon=t^{-1}_\epsilon = \epsilon^2/t_0$.
In other words, the $(z, \epsilon)$-- and 
$(t, \epsilon)$--dependence
follows from the master function $\widehat{\phi}_\lambda $ and
$\phi_\lambda$ (Eq.~(\ref{eq68})), respectively, by use of the
scaled frequency $z/\omega_\epsilon$ and time
$t/t_\epsilon$. The time scale $t_\epsilon$ diverges as
$\epsilon^{-2}$ when approaching the glass transition independent
on the sign of $\epsilon$.

\section{SUMMARY AND CONCLUSIONS}

Our main motivation has been the microscopic description of the
glass transition for systems with trivial statics, \emph{i.e.}
systems which do not undergo an equilibrium phase transition to an
ordered phase and do not have static correlations. For such
systems, present microscopic theories like MCT and replica theory
for structural glasses predict neither a dynamical nor a static
glass transition. As a model we have chosen $N$ \textit{infinitely
thin} hard rods with length $L$ fixed with their centers on a
periodic lattice with lattice constant $a$. The only relevant
physical parameter is the dimensionless length $l=L/a$.
Simulations for an fcc \cite{30} and an sc lattice \cite{31}
strongly suggested that  
a \textit{dynamical} glass transition occurs
at a critical length $l_c^{\mathrm{sim}} \cong 2.7$ \cite{30} and
$l_c^{\mathrm{sim}} \cong 4.5$ \cite{31}. Note that the rods in
ref. \cite{31} were fixed at the lattice sites with one of their
end points.

To describe the dynamics of our model we have used the generalized
$N$-rod Smoluchowski equation from which we derived a hierarchy of
coupled equations for $\delta \rho^{(k)}_{n_1 \cdots n_k}
(\Omega_1, \cdots, \Omega_k)$, the fluctuations of the reduced
$k$-rod densities. Truncating at the second level and
approximating the influence of surrounding rods on a pair of rods
by the introduction of an effective diffusion tensor
${\mathbf{D}}$ we finally obtained a self-consistency equation for
$\mathbf{D} $ (Eq.~(\ref{eq25})). This equation describes a
feedback mechanism (as MCT also does) which leads to a slowing
down of the dynamics and ultimately to an orientational glass
transition, similar to that occuring in plastic crystals \cite{5a,6,7,5,
10,11,12}. The model studied in the present paper may be applied
to real plastic crystals as has been shown in ref. \cite{5}.

One of the essential features of present MCT for the glass
transition is the cage effect \cite{13}. For low temperature or
high density an \textit{extended} particle, \emph{e.g.} a hard
sphere, is captured in a cage. The occurrence of this cage is
accompanied by the growth of \textit{static} correlations.
Particularly, the main peak of the static structure factor grows
and this leads to an increase of the static vertices entering the
MCT-equations. If the vertices reach a critical strength the
system will undergo a \textit{dynamical} glass transition,
provided ordering is prevented. In this sense the cage effect
entering into MCT through the static correlations is of
\textit{static} nature. The model studied here does not
have any static correlations, yet there is a cage effect. If the
rods are long enough the tagged rod's motion can be strongly
restricted to an ``orientational'' cage (cf. Figure 2). This cage,
however, is of pure \textit{dynamical} nature and leads to a glass
transition as well.

Comparing the results for the fcc lattice obtained from our
theoretical framework with the additional approximation
Eq.~(\ref{eq31}) with those from simulations \cite{30,31} one can
say that both values $l_c^{\mathrm{num}}\cong 2.20$ and
$l_c^{\mathrm{var}}\cong 3.96$ are in a reasonable range of
$l_c^{\mathrm{sim}}\cong 2.7$ \cite{30}. Taking into account that
our approach involves uncontrollable approximations
(Eqs.~(\ref{eq19}) and ~(\ref{eq20})) this is a satisfactory
result. The tendency of $l_c$ to decrease when going from the 
sc lattice to the bcc to the fcc lattice is consistent with our
expectation that $l_c$ decreases with increase of the coordination
number and decrease of the nearest neighbor distances. Of course,
second-, third-, \textit{etc.} nearest neighbors play also a role, 
but with minor influence. Less satisfactory is the exponent, 
$\gamma$ of the power law,
$D(l) \sim (l_c - l)^\gamma$, which is $\gamma=1$, in contrast to
$\gamma^{\mathrm{sim}} \cong 4.2$ \cite{31}. Even if the simulated
result turns out too high, the type-A transition result, 
$\gamma=1$, cannot be the correct value. This
requires an improvement of our microscopic approach.
It is not obvious whether the
approach presented in ref. \cite{23} which is based on rather
qualitative arguments can serve as a guide.

It seems that the ``dynamical cage effect'' leads to a
discontinuous variation of the nonergodicity parameter at $l_c$
and is accompanied close to $l_c$ by a two-step relaxation
process, as can be seen from the orientational correlator studied
in ref. \cite{30}. Such a two-step relaxation was first predicted
by MCT \cite{1,13,14,15} and within this theory it is related to
the cage effect. The present theoretical approach predicts a
continuous transition for the glass order parameters $f_\lambda$.
Therefore it does not yield a two-step relaxation. This fact
demands for an improvement of our theory, based on the hierarchy
of equations for the $k$-rod densities. But on the other hand it is,
of course, a big challenge to extend the present MCT such that it
includes the \textit{``dynamical cage effect''} described above
and that a discontinuous glass transition occurs. If this turns
out to be possible, it seems that the vertices must be
generalized. Besides static correlations, \emph{i.e.} correlators
\textit{at time $t=0$}, they also must contain dynamical ones.
These dynamical correlations may enter through a time-dependent
force-force correlator $\int\limits_0^{\infty} d t \langle
\mathbf{F} (t) \cdot \mathbf{F} (0) \rangle$, \emph{i.e.} as a
Laplace transformed correlator \textit{at frequency $z=0$} (as
found in the present approach). For systems with vanishing static
correlations, like for our model, the vertices would be of purely
dynamical nature. Now, increasing $d$ to finite values will
generate static correlations. For small thickness the ``dynamical
cage'' will be still dominant, but at a crossover value
$d_{\mathrm{c.o.}}$ the static cage effect would become comparable
with the dynamical one and for $d > d_{\mathrm{c.o.}}$ it would be
the dominant one.

Let us come back to the choice of Brownian {\it vs.} Newtonian
dynamics. We have found that the glass transition is driven by the
increase of the Laplace transform at $z=0$ of the time dependent 
torque-torque correlator of an isolated pair of rods. 
This correlator is be different for Brownian and
Newtonian dynamics. Therefore the critical length $l_c$ will be not the
same for both dynamics. At present, it is not clear whether this is
only a small effect. This is different from what has been found 
for liquids with nontrivial static correlations \cite{34,35}.
However, the linear dependence of the nonergodicity parameters and 
of the diffusion constant on $\epsilon$, the quadratic dependence of the 
frequency scale $\omega_{\epsilon}$ on $\epsilon$ and
the $t^{-1/2}$ dependence of the correlator at the critical length $l_c$ 
should be independent of the type of dynamics.

It would be interesting to re-investigate the lattice model with
infinitely thin hard rods by computer simulations. With present
computers it should be possible to cover a larger time range and
to study the dynamics close to $l_c$ in greater detail. This would
allow to check whether the two scaling laws and other
predictions of MCT are consistent with simulations. 
If the outcome of such a test proves
consistency it would encourage theoretical efforts to extend the
microscopic theory within the framework of MCT.

Besides rods on a lattice one could also investigate liquid
systems of ``particles'' built by crossing infinitely thin hard
rods. If the number of ``legs'' (rods) of a ``particle'' becomes
large they may exhibit quite similar dynamical behavior to hard
spheres, although there are no static correlations. The time scale
on which these ``particles'' realize that they are not hard
spheres could be much larger than a typical time scale for
structural relaxation.

To conclude, we have discussed a \textit{purely dynamical}
mechanism which also drives a glass transition and which up to now
has received almost no attention. Its further investigation by
computer simulations and analytical work seems to us a challenging
task for the future.
\bigskip
\bigskip

{\bf{ACKNOWLEDGMENT}}
\bigskip

We gratefully acknowledge a very fruitful and stimulating
discussion with W. G\"otze during the ``3$^\mathrm{rd}$ Workshop
on Non-Equilibrium Phenomena in Supercooled Fluids, Glasses, and
Amorphous Materials" in Pisa and his comments 
on the present manuscript. Helpful comments by M. Fuchs and the
preparation of most of the figures by D. Garanin and M. Ricker are
gratefully acknowledged as well. G. S. acknowledges support from
the National Science Foundation through grant No. CHE 0111152 and
the Alexander von Humboldt Foundation.

\newpage

\begin{figure}
\caption{\label{fig1}
Unit cell of the sc-lattice with lattice constant $a$ with hard
rods of length $L$.}
\end{figure}

\begin{figure}
\caption{\label{fig2} Illustration of the blocking of rod 1 by
rods 2-5 within shaded region for $l>2$..}
\end{figure}

\begin{figure} \caption{\label{fig3}
Illustration of possible collisions of rod $i$ and $j$ for fixed
rod $i$. For better visualization a small but finite thickness has
been used. (a) Single collision for $l<2$ at
$\phi_j=\phi_i^{\pm}$. The rod $j'$ (dashed) turned around z-axis
by $\pi$ does not lead to a second collision if $l<2$. (b) Single
collision for $l<2$ at $\phi_j=\phi_i^{\pm} + \pi$. Again a
rotation by $\pi$ (not shown) does not yield a second collision
(c) Two collisions for $l>2$; $\theta_i$ and $\theta_j$ are chosen
such that two collisions occur at $\phi_j=\phi_i ^{\pm}$ and
$\phi_j=\phi_i^{\pm} + \pi$.}
\end{figure}

\begin{figure} \caption{\label{fig4}
Illustration of geometrical quantities defined in the text for
two rods $i$ and $j$.}
\end{figure}

\begin{figure} \caption{\label{fig5}
Time-dependence of the torque-torque correlator of an isolated
pair of rods for different $l/r_n$. Dotted line: $l/r_n=1.8$;
dash-dotted line:
$l/r_n=2$; dashed line: $l/r_n=4$; solid line: $l/r_n=6$}
\end{figure}

\begin{figure} \caption{\label{fig6}
$\upsilon_0(x)$ from the numerically exact (squares) and the
variational calculation (solid line).}
\end{figure}

\begin{figure} \caption{\label{fig7}
$\upsilon(l)$ for a sc-lattice from the numerically exact
(squares) and the variational calculation (solid line). $l_c$ denotes
the
critical length for which $\upsilon(l_c)=1$.}
\end{figure}


\begin{thebibliography}{99}
\bibitem{43}
Our numerical result (Fig.~\ref{fig5}) is consistent
with this assumption. It is possible that the decay is
exponential due to the compact configuration space of two
needles in contrast to two hard spheres.
\end{thebibliography}
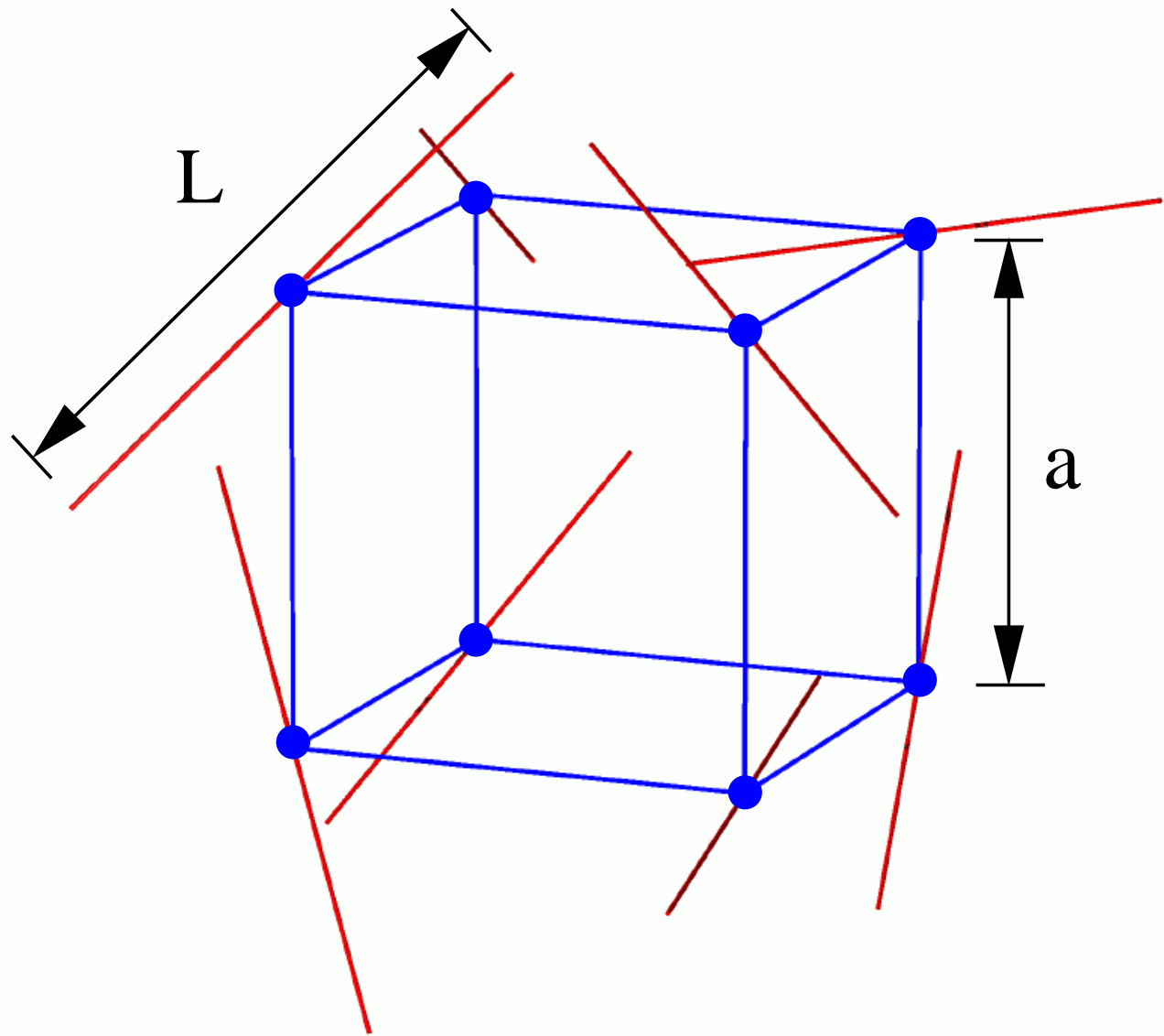
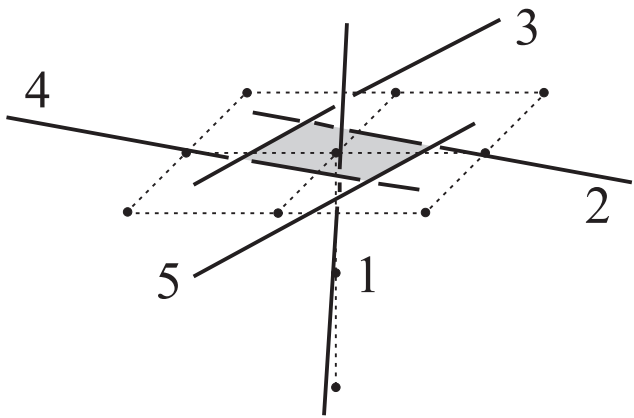
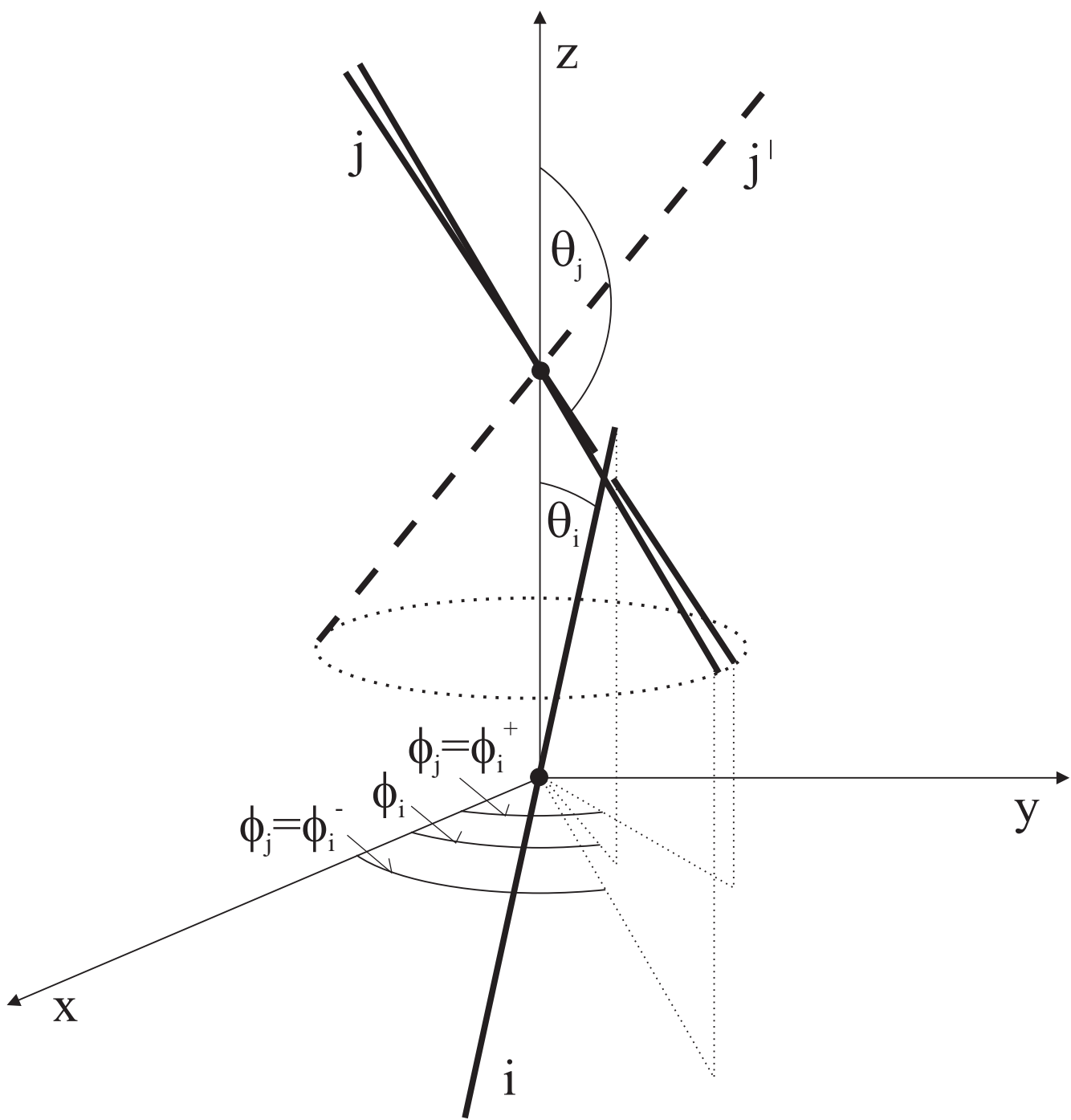
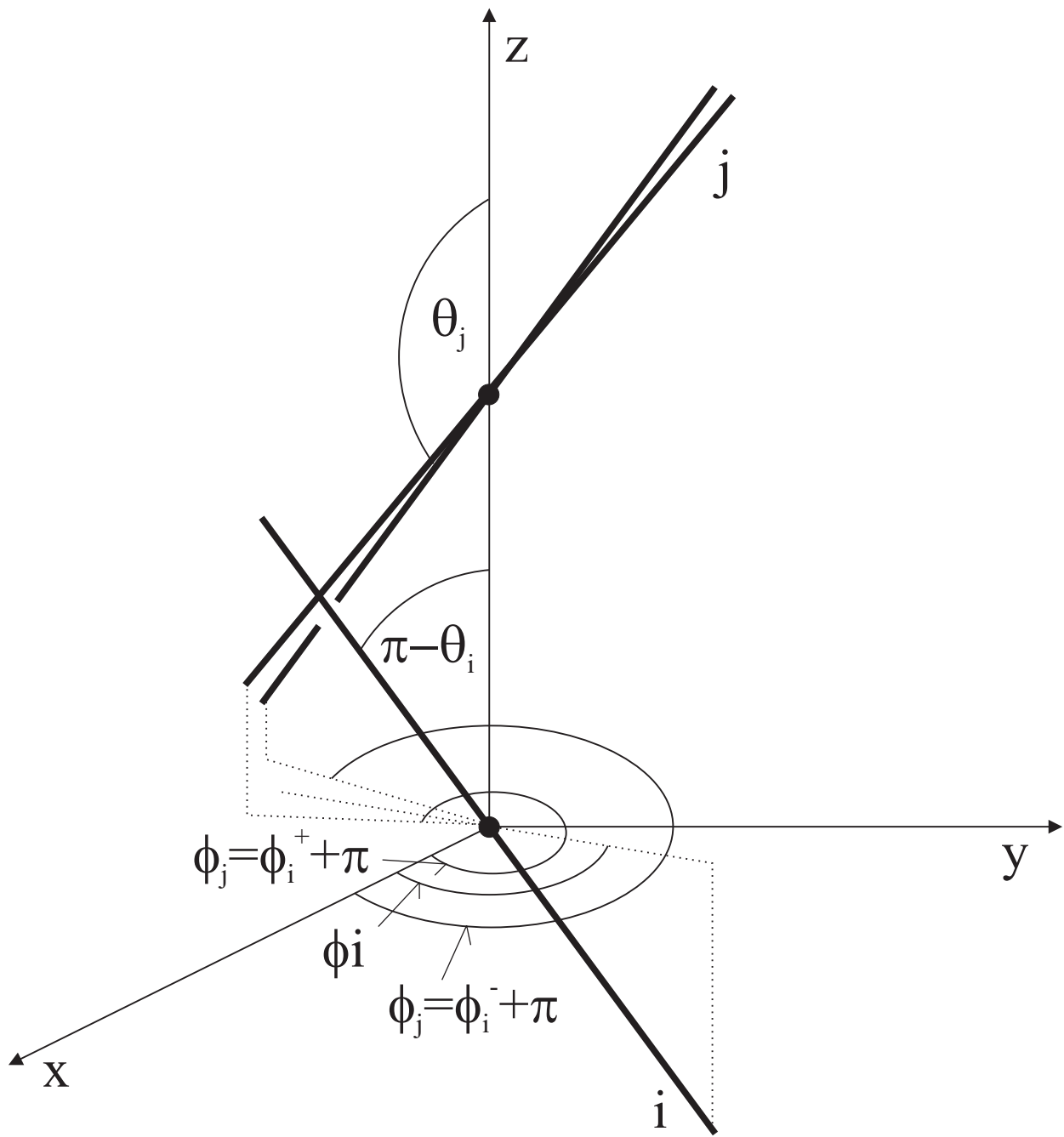
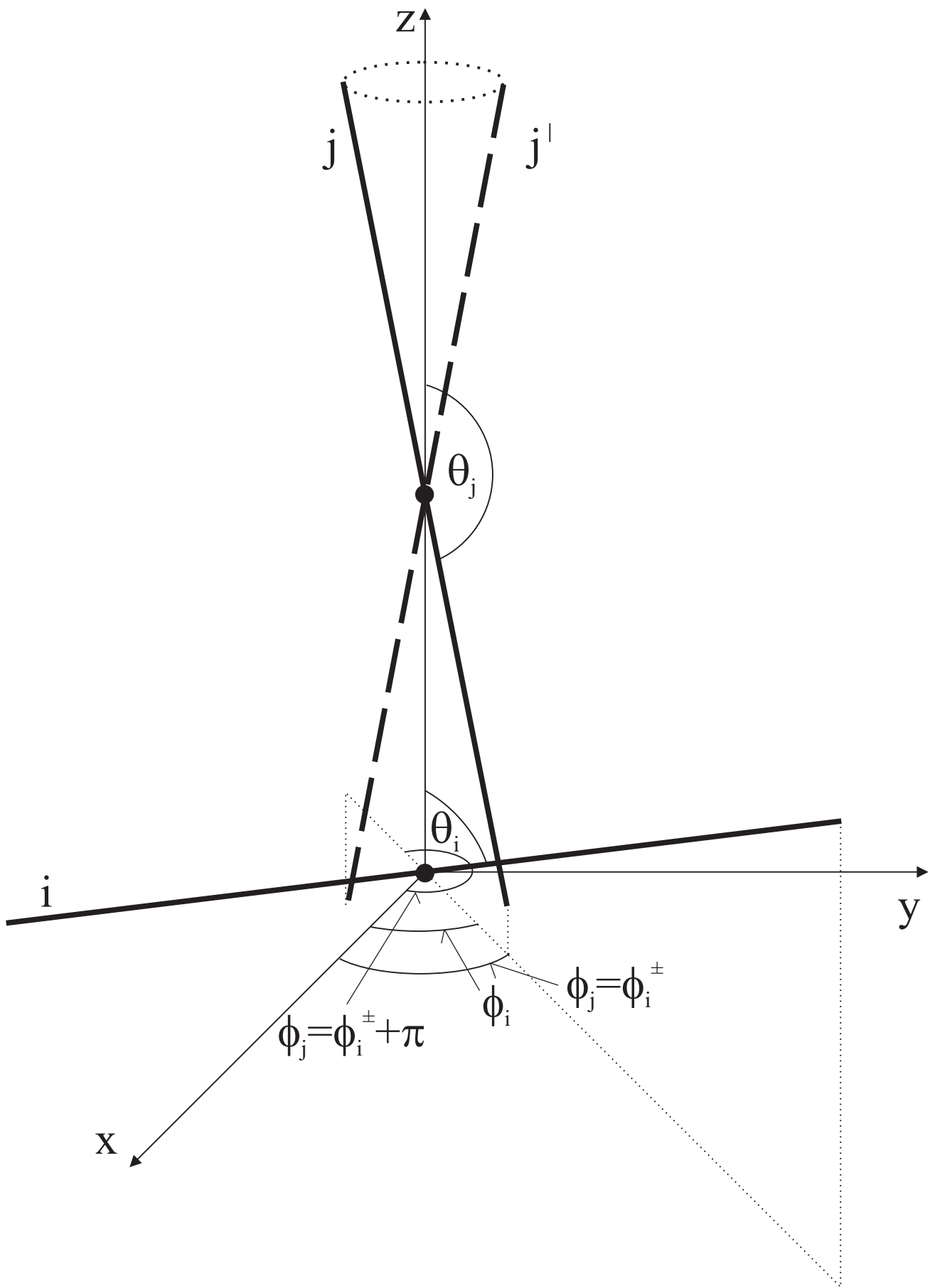
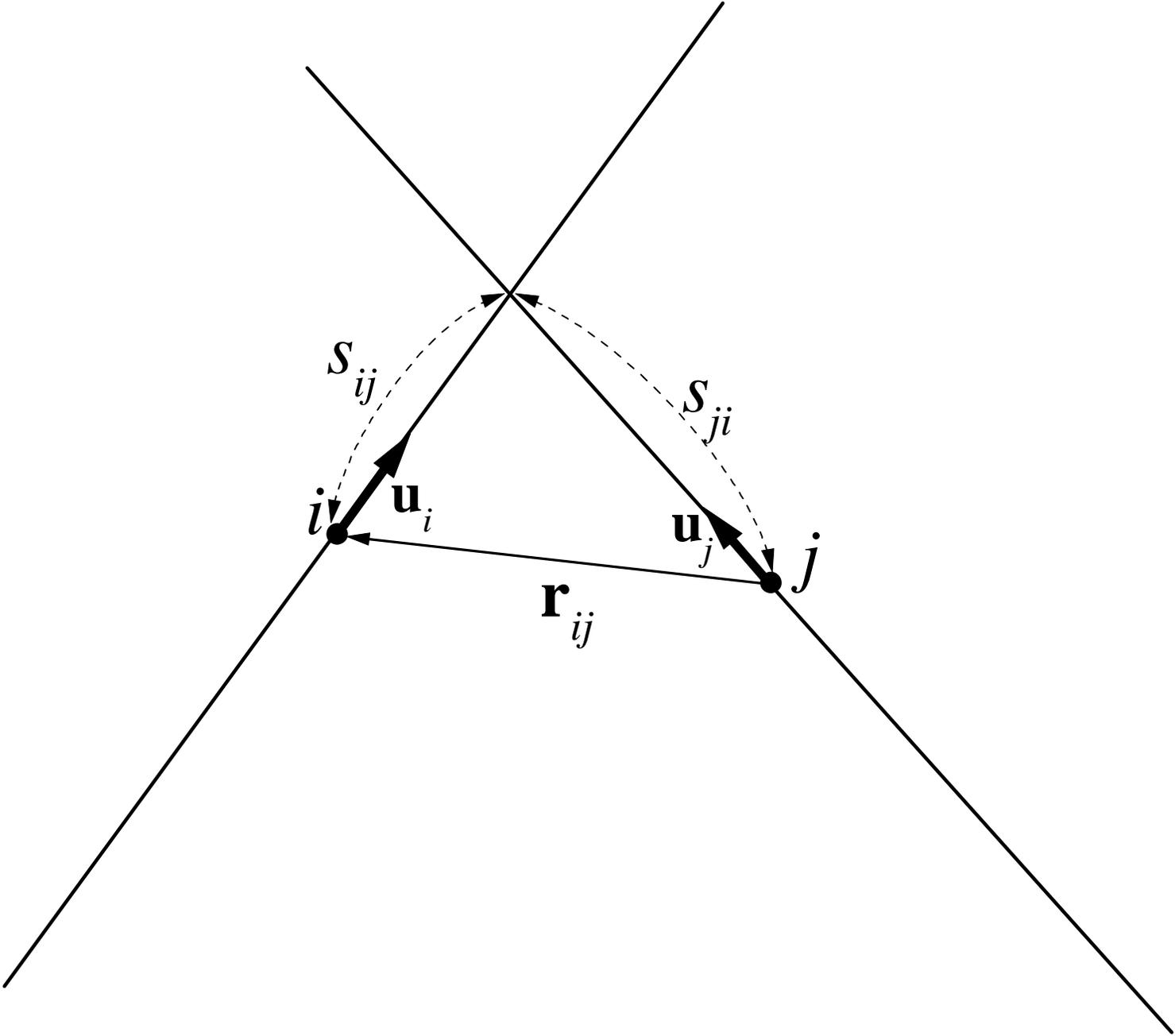
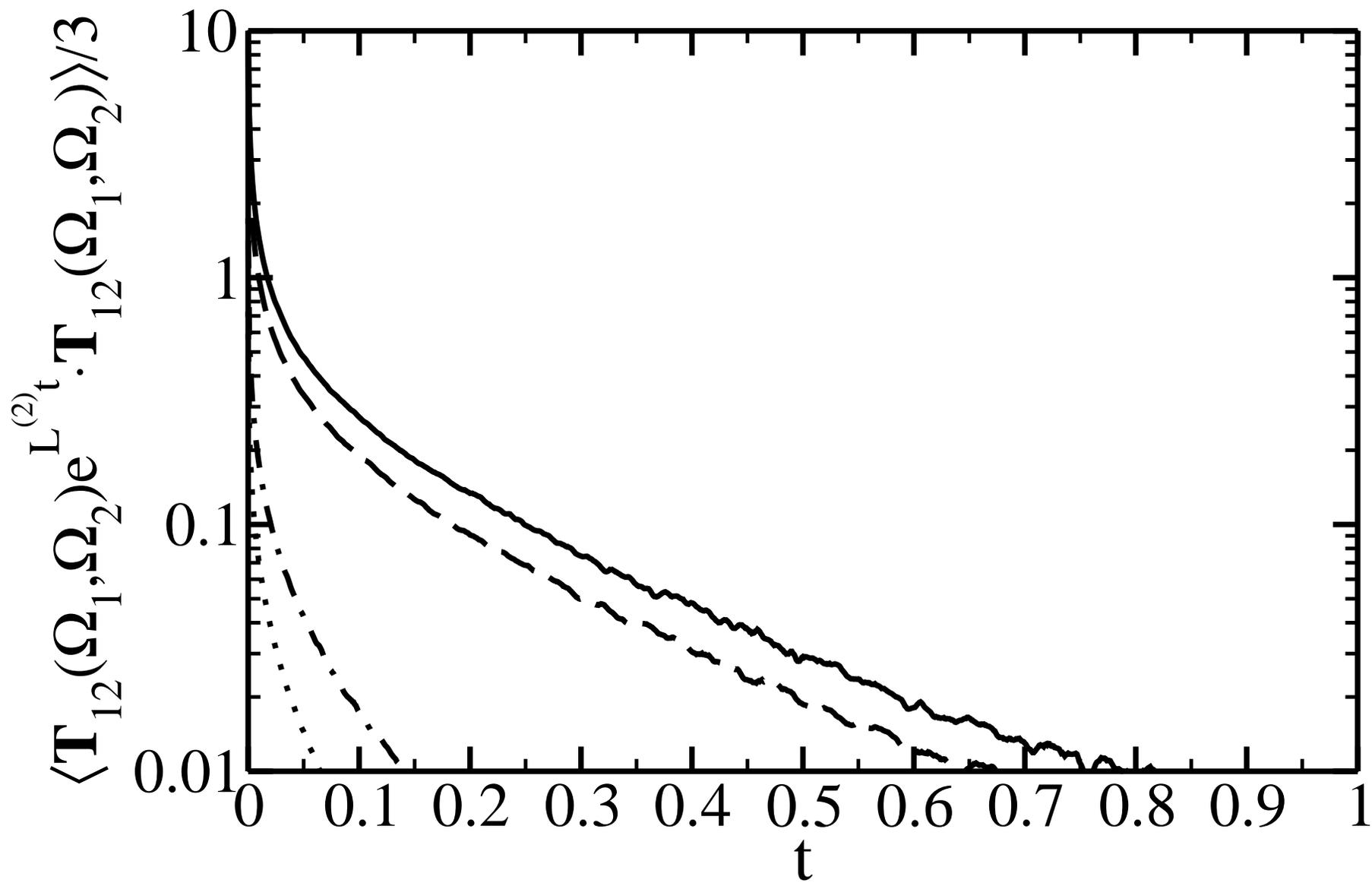
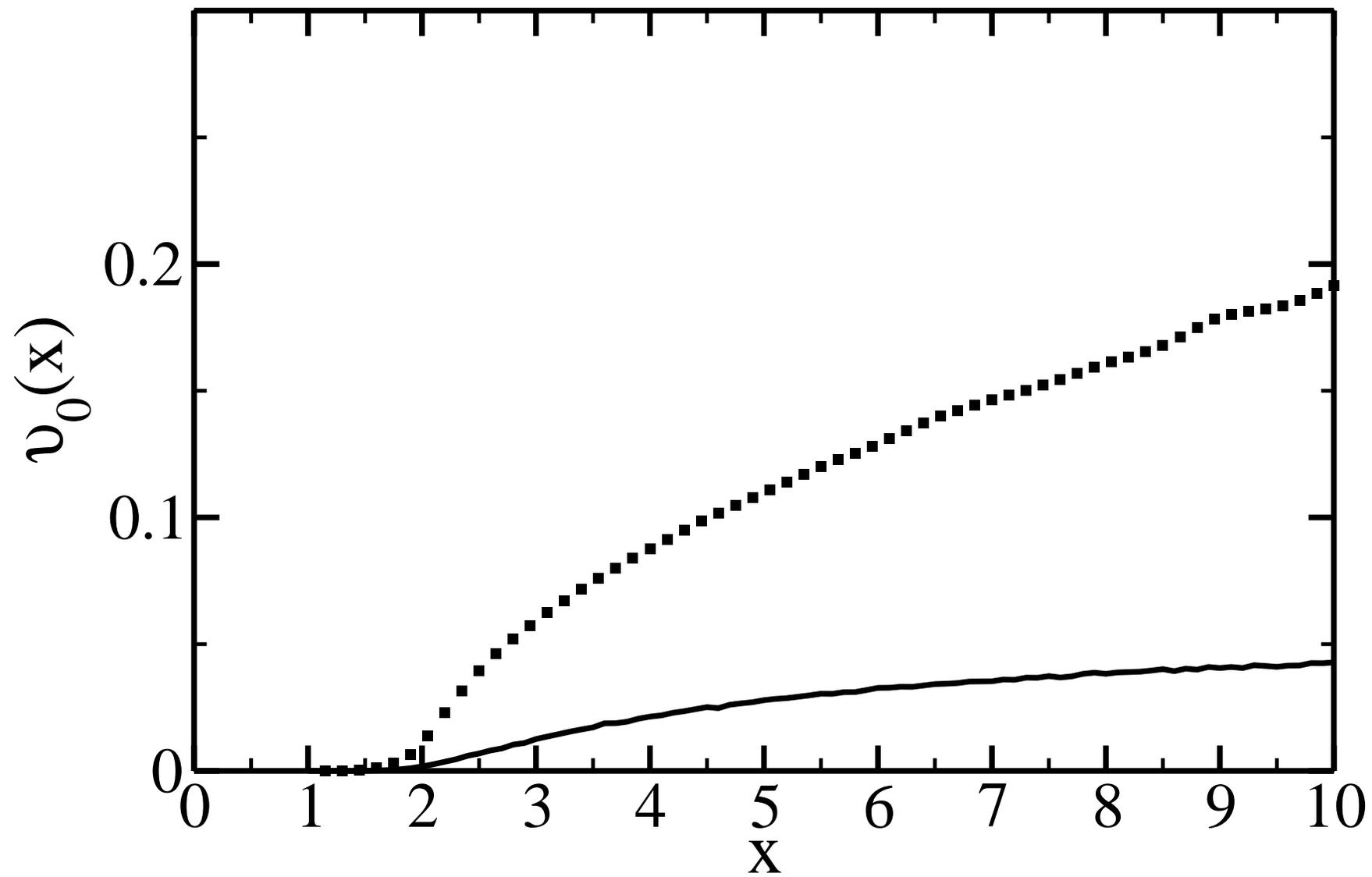
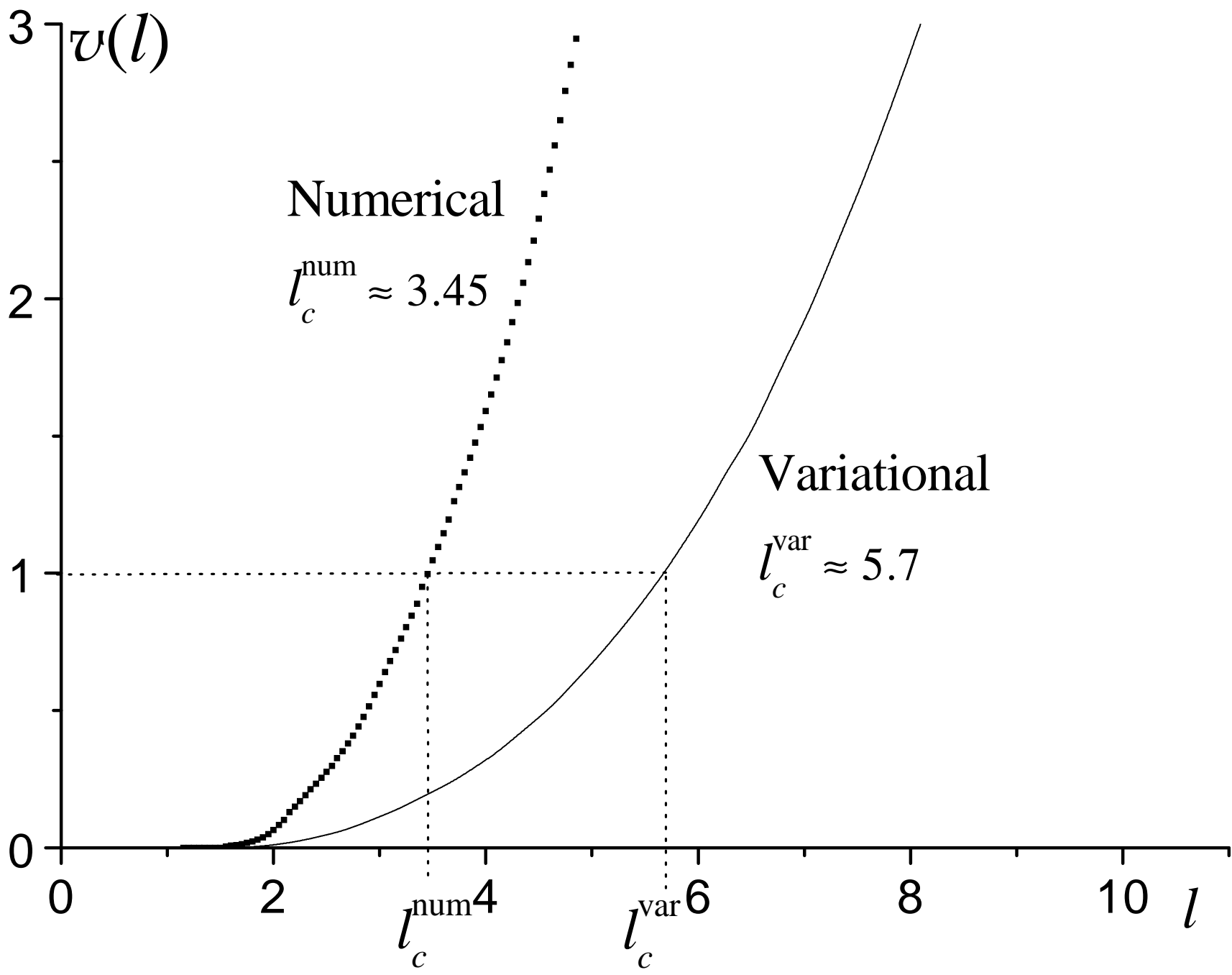
\end{document}